\documentclass[twocolumn]{aastex63}
\usepackage{array}
\usepackage{multirow}
\usepackage{xcolor} 
\usepackage{amsmath}

\newcolumntype{L}{>{\centering\arraybackslash}m{3cm}}
\newcolumntype{M}{>{\centering\arraybackslash}m{1cm}}

\received{August 03, 2020}
\revised{November 4, 2020}
\accepted{November 8, 2020}

\submitjournal{ApJ}

\shorttitle{Small-Scale Dynamo in Stably Stratified Turbulence}
\shortauthors{V. Skoutnev, J. Squire, A. Bhattacharjee}
\graphicspath{{./}{}}

\begin{document}

\title{Small-Scale Dynamo in Stably Stratified Turbulence}

\correspondingauthor{Valentin Skoutnev}
\email{skoutnev@princeton.edu}

\author{V. Skoutnev}
 \affiliation{Department of Astrophysical Sciences and Max Planck Princeton Center, Princeton University, Princeton, NJ 08544, USA}
\author{J. Squire}
\affiliation{Physics Department, University of Otago, Dunedin 9010, New Zealand}
\author{A. Bhattacharjee}
\affiliation{Department of Astrophysical Sciences, Princeton University, Princeton, NJ 08544, USA}



\begin{abstract}
We present numerical investigations into three principal properties of the small-scale dynamo in stably stratified turbulence: the onset criterion, the growth rate, and the nature of the magnetic field anisotropy in the kinematic regime. The results suggest that all three dynamo properties are controlled by the scale separation between the Ozmidov scale and the viscous or resistive scale. In addition to the critical magnetic Reynolds number, this allows for the definition of critical buoyancy and magnetic buoyancy Reynolds numbers for stratified small-scale dynamo onset in the high and low magnetic Prandtl number regimes, respectively. The presence of a small-scale dynamo in stellar radiative zones could affect dynamics through resulting Maxwell stresses and/or influence large-scale dynamo mechanisms in regions of differential rotation. Taking the solar radiative zone as a representative example and applying the onset criterion, we find that the stratification is strong enough to make the small-scale dynamo marginally active in the stably stratified turbulence of the solar tachocline.
\end{abstract}



\section{Introduction} \label{sec:intro}

Magnetic fields play critical roles throughout many stages of stellar evolution. In particular, dynamo-generated-magnetic fields in radiative zones (regions of stable stratification) are thought to be able to efficiently provide torques that maintain nearly uniform rotation profiles \citep{aerts2019angular}. A leading candidate is the Tayler-Spruit dynamo \citep{spruit2002dynamo} driven by instability of a toroidal field wound up by differential rotation in a spherical geometry. Another possible candidate is the magnetorotational instability \citep{kagan2014MRI,wheeler2015role,rudiger2015angular}, a local instability based on a negative gradient in the angular velocity. However, sufficiently strong stratification in some stages of evolution (e.g. from steep composition gradients) is able to inhibit both dynamo mechanisms and suppress angular momentum transport (but see \cite{fuller2019slowing}). Magnetic fields thus appear to be fairly common in stably stratified regions, with potentially important influences on a variety of physical processes. However, the influence of stable stratification on the growth of magnetic fields in turbulence – the small-scale dynamo (SSD)  instability -- has not (to our knowledge) been previously investigated. It is thus the purpose of this paper to examine how stable stratification influences the SSD's onset, growth rate, and structure.

The SSD is typically found to accompany any dynamo mechanism due to its operation on the smallest length scales, and correspondingly fastest timescales. As a result, the SSD may complement, coexist, or compete with other present dynamo mechanisms \citep{kulsrud1992spectrum,schekochihin2002spectra}. In stably stratified regions, the SSD can be driven by, in principle, turbulence generated by horizontal/vertical shear instabilities, breaking internal gravity waves, and/or convective overshoot. An unstable SSD should saturate with rough equipartition between magnetic field energy and turbulent kinetic energy, which could have two important effects. First, Lorentz forces become strong enough to feed back on the fluid turbulence and, in a region of differential rotation, could supply Maxwell stresses that contribute to angular momentum transport. Second, background, fluctuating small-scale magnetic fields are also known to significantly influence any operating large-scale dynamo through quenching \citep{vainshtein1992nonlinear,gruzinov1994self,bhattacharjee1995self,zhouGenQuench}, helicity fluxes \citep{blackman2000constraints,vishniac2001magnetic,ebrahimi2014helicity}, and magnetic shear-current effects \citep{squirePRL_MSC}. The net effect of the small-scale dynamo on large-scale field growth is therefore not immediately obvious. Understanding the interplay of these effects in a realistic astrophysical setting is a difficult task; as a first step, it is important to investigate the instability criterion, growth rate, and magnetic field structure of the stably stratified SSD.

\subsection{Small-Scale Dynamos}\label{sec:IntroSSD}
The SSD has been extensively studied in the unstratified case, which we briefly review. The SSD is categorized as growth and sustenance of magnetic fields on length scales $l$ smaller than the turbulent integral (forcing) scale $l_i$ in a conducting fluid, differentiating itself from the large-scale dynamo which grows on scales $l>l_i$ due to some broken symmetry in the turbulence such as shear or helicity \citep{brandenburg2005astrophysical}. A sufficient initial condition for the SSD to begin operating is a local, random, weak seed field, an astrophysical requirement often easily satisfied. In realistic astrophysical systems, the SSD operates either in the high $Pr_m\gg1$ (interstellar medium or outer regions of accretion disks) or low $Pr_m\ll1$ (stellar and planetary interiors or inner regions of accretion disks) regimes, where the magnetic Prandtl number $Pr_m=\nu/\eta$ is the ratio of the fluid viscosity and magnetic resistivity. Its behavior, stability, and growth rates can depend strongly on $Pr_m$. 

The high-$Pr_m$ regime has been extensively studied analytically and numerically because the much smaller size of the resistive scale $l_\eta\sim Pr_m^{-1/2}l_\nu$ compared to the viscous scale $l_\nu$ allows for the viscous-scale velocity field acting on the magnetic field to be modeled as a random and spatially smooth viscous flow \citep{kazantsev1968enhancement,zel1984kinematic, Schekochihin2004}. The dynamo-generated fields are characterized by folds that are straight up to the scale of the flow with field-direction reversals on resistive scales and a growth rate comparable to the turnover time scale of the viscous eddies. 

On the other hand, in the low $Pr_m$ regime the resistive scale $l_\eta\sim Pr_m^{-3/4}l_\nu$ sits inside the inertial range where the lack of time and length scale separation between magnetic field stretching and diffusion makes dynamo action difficult to model. Numerical simulations strongly suggest its existence and show that its critical magnetic Reynolds number $Rm^c$, the $Rm=Pr_m Re$ above which the SSD turns on, is much larger than in the high $Pr_m$ regime but still reaches a finite limit for $Pr_m\rightarrow 0$ \citep{Iskakov2007, Schekochihin2007}. This is in qualitative agreement with the analytical model of \cite{BoldCat}, which predicts a higher $Rm^c$ at low $Pr_m$ due to the rougher velocity spectra in the inertial range compared to the viscous range.  

\subsection{Addition of Stratification}
The assumptions of isotropy and homogeneity of the background turbulence are typically used for drastic theoretical and computational simplification in SSD theory. They are often a good approximation in subregions of many large systems until fields become strong enough to anisotropically feed back onto the fluid flow starting from the smallest scales, eventually saturating the dynamo \citep{Schekochihin2004a}. However, in the context of stellar interiors, these assumptions break down at large-scales in regions of shear flows, convection, and, our focus, stable stratification. Numerical studies of convection find robust SSD growth near unity magnetic Prandtl number \citep{Graham_2010,favier2011small,Hotta_2015,Yadav_2015,borrero2017solar}. At a more realistic lower value of $Pr_m=0.1$, the highest-resolution $Rm\approx 100$ convection simulations have yet to demonstrate a positive SSD growth \citep{kapyla2018small}, due to the significantly increased computational cost and a potentially even larger $Rm^c>300$ than in the isotropic-forcing case of \cite{Schekochihin2007} and \cite{Iskakov2007} at the same $Pr_m$. However, stellar convection zones at $Rm=O(10^{13})$ are easily above $Rm^c=O(10^2)$, and therefore the SSD is expected to be universal in convective turbulence \citep{borrero2017solar}. On the other hand, the effect of stable stratification on the SSD has not been examined to our knowledge.

Stable stratification generates significant anisotropy by restricting vertical fluid motions in favor of horizontal fluid motions, modifying the growth rate and saturation of the SSD. It is well known that too much anisotropy will shut off the SSD. Indeed, it can be proven that a two-component, three-dimensional velocity field cannot sustain a dynamo \citep{zel1984kinematic}. This begs the first important question this paper attempts to answer: what is the dynamo onset criterion in the presence of stratification? Numerical investigations in Section \ref{sec:Simulations} suggest that the modified dynamo onset criterion is, in addition to $Rm^c$, set by a critical buoyancy Reynolds number $Rb^c$ for high $Pr_m$ and a critical magnetic buoyancy Reynolds number $Rb_m^c$ for low $Pr_m$, where we define $Rb_m=Pr_m Rb$. A physical understanding of this criterion is discussed in Section \ref{sec:LengthScales}. When the onset criterion is satisfied, the second question naturally follows: what is the anisotropy of the dynamo-generated magnetic field in the kinematic limit? Spectral diagnostics in Section \ref{sec:Simulations} find that the anisotropy in the magnetic field is primarily set by the anisotropy of the velocity field at the viscous/resistive scales for the high/low $Pr_m$ regimes. Following the kinematic regime, the dynamo will eventually saturate. We leave understanding the properties of the saturated field for future study. 

In application, we extrapolate our results to the Sun as a representative of main-sequence stars and consider the solar tachocline, for which helioseismology and solar models provide parameter estimates. Stratified turbulence in the tachocline is thought to be driven by a combination of overshoot from the overlying solar convection zone and shear instabilities sourced by solar differential rotation across and along the layer \citep{miesch2005large}. With the resulting large kinetic and magnetic Reynolds numbers typically calculated for the region, an estimate of the SSD growth rate that neglects stratification suggests the SSD would be very active. However, we find that the stratified SSD onset criterion is only marginally satisfied, highlighting the importance of considering the effects of stratification on the SSD (see Section \ref{sec:ApplicationToTC}). This suggests that while equipartition small-scale magnetic fields may be present in the tachocline (absorbing energy from the stratified turbulence, providing additional Maxwell stresses, and influencing any operating large-scale dynamo mechanism), the SSD may be suppressed in other parts of the solar radiative zone where driving mechanisms for stratified turbulence are expected to be weaker. Generalizing to other stars, we predict that the SSD may significantly vary in strength depending on the local level of differential rotation, similar to other radiative-zone dynamo mechanisms.

\subsection{Paper Outline} 
Section \ref{sec:LengthScales} presents an overview of stratified turbulence by examining the energy cascade and important length scales. Section \ref{sec:InterpretationOfSims} combines all the simulation growth rates to present the suggested dynamo onset criterion. Section \ref{sec:ThermalPr} discusses the role of the thermal Prandtl number.

Section \ref{sec:Simulations} presents the direct numerical simulations in detail. Section \ref{sec:setup} describes setup of the simulations and Section \ref{sec:Diagnostics} defines spectral diagnostics used for analysis. Section \ref{sec:Results} then presents results obtained for the $Pr_m=1$, high $Pr_m$, and low $Pr_m$ regimes. 

Section \ref{sec:ApplicationToTC} discusses application to stellar radiative zones. Section \ref{sec:conc} summarizes and concludes.

\section{Phenomenology of Stably Stratified Turbulence}\label{sec:LengthScales}
We review a phenomenological picture of the energy cascade across several length scales of stratified turbulence. An alternative, but closely related perspective of stably stratified turbulence via a scaling analysis of the governing Boussinesq equations and its extension to the magnetic induction equation is presented in Appendix \ref{sec:ScalingAnalysis}.

\subsection{Energy Cascade and Length Scales}
\begin{figure}
    \centering
    \includegraphics[width=\linewidth]{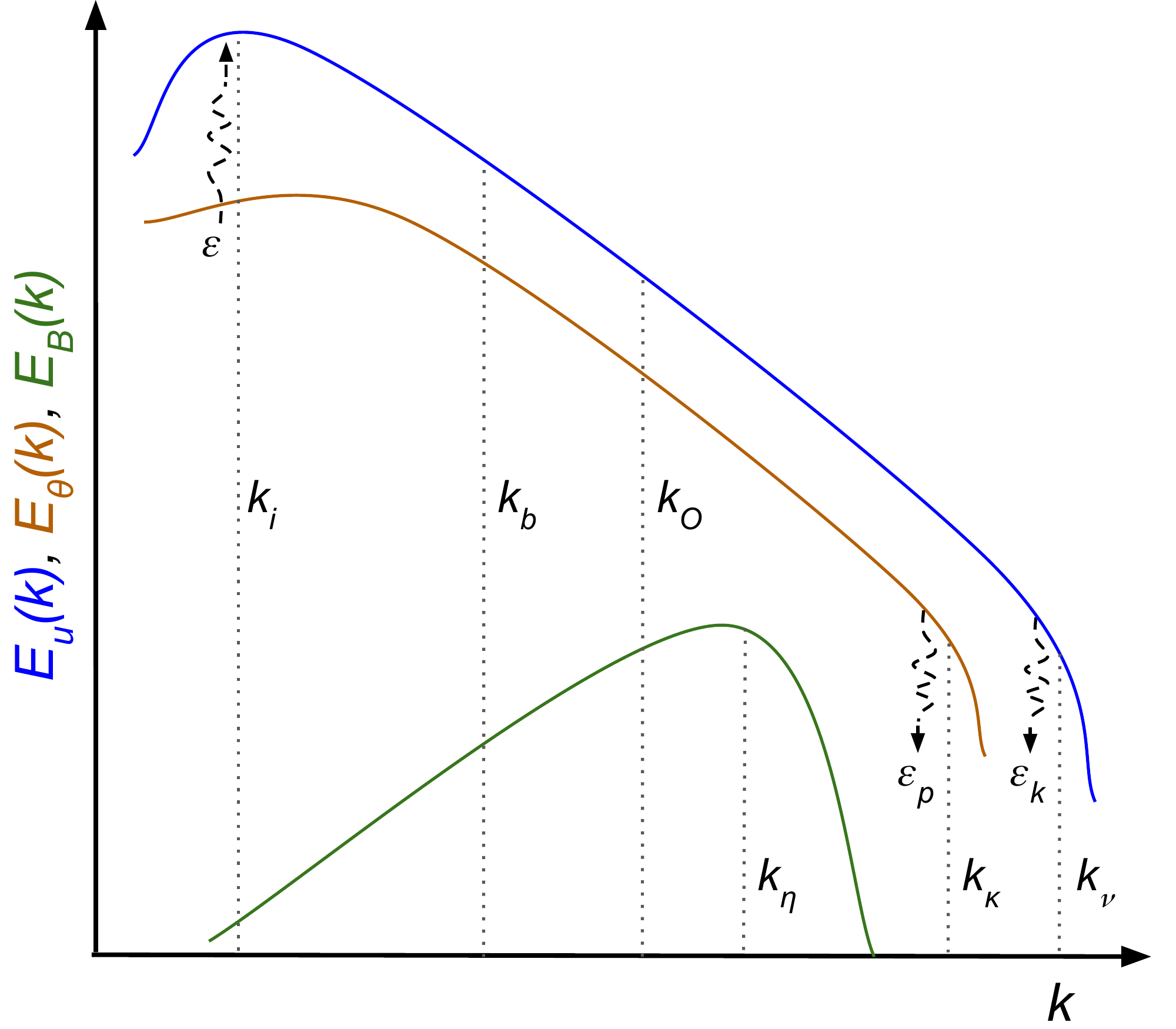}
    \caption{Sketch of kinetic $E_u(k)$, buoyancy $E_\theta(k)$, and magnetic $E_B(k)$ energy spectra for $Pr_m<Pr<1$. All parameters are defined in Section \ref{sec:LengthScales}. }
    \label{fig:SpectraSketch}
\end{figure}

Stratified turbulence can be understood by examining the energy cascade, whose anisotropy is strongly scale dependent. We consider only the kinematic limit where energies in the magnetic fields are too small to affect the fluid motion and the standard hydrodynamic picture holds. Kinetic energy injected at a rate $\epsilon$ at the integral scale $l_i$ is dissipated through viscous, $\epsilon_k$, thermal, $\epsilon_p$, and resistive, $\epsilon_m$ ($\epsilon_m\ll \epsilon$), dissipation channels ($\epsilon=\epsilon_k+\epsilon_p+\epsilon_m$). The ratio $\epsilon_p/\epsilon_k$ is determined by the Froude number $Fr=u_{\rm rms}/(Nl_i)$ and approaches quasi-equipartition $\epsilon_p\lesssim \epsilon_k$ at low-enough $Fr$ \citep{lindborg_2006,pouquet_2018}, where $u_{\rm rms}$ is the root-mean-square fluid velocity and $N>0$ is the Brunt-V\"ais\"al\"a frequency (for definition, see Appendix \ref{sec:ScalingAnalysis}). Unlike Kolmogorov turbulence, which has a single inertial range, stratified turbulence exhibits three distinct ranges whose scale separations are controlled by $Fr<1$ and $Re=u_{\rm rms}l_i/(2\pi\nu)$. At large-scales, instabilities in a stratified fluid with no vertical variation (such as the zigzag instabilities; \citealp{billant2000theoretical}) restrict vertical scales to below the buoyancy length $l_b=u_{\rm rms}/N=Fr l_i$ (alternatively, the scale above which gravity restricts eddies from turning over in the vertical direction). In a fluid with $l_i > l_b$, large-scale turbulence is dominated by pancake vortices and internal gravity waves that can transfer energy directly to the buoyancy scale through Kelvin-Helmholtz instabilities of vertically adjacent vortices and overturning of internal gravity waves \citep{Waite2011,carnevale_briscolini_orlandi_2001,waite_bartello_2006}. The energy brought to the buoyancy scale is then transferred through an anisotropic cascade down to the Ozmidov scale $l_{O}=(\epsilon/N^3)^{1/2}$ where the local eddy turnover frequency matches the Brunt-V\"ais\"al\"a frequency $N$. For smaller scales $l<l_O$ inertia dominates gravity, so the Ozmidov scale acts as an outer scale for a quasi-isotropic Kolmogorov cascade down to the viscous scale $l_\nu\sim Rb^{-3/4}l_O$, where $Rb$ is the buoyancy Reynolds number (see below). Thermal energy is likewise removed at the thermal dissipation scale $l_\kappa\sim l_\nu$ when the thermal Prandtl number is order unity, $Pr=\nu/\kappa\sim1$.

In summary, defining wavenumbers $k=2\pi/l$ corresponding to scales $l$, the scale separations relative to $k_i$ are given by 
\begin{equation}
k_i:k_b:k_O:k_\nu,
\end{equation}
\begin{equation}
1:Fr^{-1}:Fr^{-\frac{3}{2}}:Re^{\frac{3}{4}},
\end{equation}
with $k_\eta\lessgtr k_\nu$ depending on whether $Pr_m\lessgtr1$. A sketch of the energy spectra and relative locations of wavenumbers is shown in Figure \ref{fig:SpectraSketch}. 

The scale separation between the stratification scales and the viscous scale determines the nature of the turbulence. In particular, the ratio of the Ozmidov scale to the viscous scale $k_\nu/k_O=Rb^{3/4}$ is dependent on the buoyancy Reynolds number $Rb=ReFr^2$ and has been found to be the relevant parameter determining the transition between two regimes of stratified turbulence (see Section \ref{sec:ScalingAnalysis} for further detail). When $Rb>1$, simulations typically exhibit large horizontal layers (pancake vortices) in the presence of Kelvin-Helmholtz-type vortices, internal gravity waves, and smaller-scale 3D turbulent-like structures \citep{lindborg_2006,brethouwer_billant_lindborg_chomaz_2007,Waite2011}. This is known as the stratified turbulence regime. When $Rb<1$, simulations are typically characterized by thin, large-scale, stable horizontal layers that are missing smaller-scale features due to the suppression of instabilities and the transition to turbulence by viscosity \citep{brethouwer_billant_lindborg_chomaz_2007}. This is known as the viscosity-affected stratified flow regime (VASF). In summary, a large quasi-isotropic range $k_O\ll k_\nu$ corresponds to the strongly stratified turbulence regime ($Rb\gg 1$) while a highly viscous or too strongly stratified fluid leads to the VASF regime when $k_O>k_\nu$ ($Rb<1$).

Returning to kinematic dynamo theory, the obvious question is, how do these stratification scales relate to the dynamo growth rate at high and low $Pr_m$? A priori, one would expect the highly anisotropic eddies at the largest scales ($k\lesssim k_b$) would not contribute to the dynamo, while eddies in the quasi-isotropic subrange ($k\gtrsim k_O$) would. The contribution of the buoyancy subrange $k_b<k<k_O$ is then a priori uncertain. For high $Pr_m$, the fluid viscous scale eddies $k\sim k_\nu<k_\eta$ primarily set the SSD growth rate, so the high $Pr_m$ dynamo could potentially survive into the VASF regime when $k_b<k_\nu< k_O$ (i.e. when $Rb<1$). For low $Pr_m$, the fluid resistive scale eddies $k\sim k_\eta<k_\nu$ are thought to set the SSD growth rate, and the question becomes whether the low $Pr_m$ SSD can survive in an increasingly stratified regime when $k_b<k_\eta< k_O<k_\nu$ (i.e. when $Rb_m<1$). 

\section{Interpretation of Simulations}\label{sec:InterpretationOfSims}
In this section, we combine the results of the direct numerical simulations (DNSs) in Section \ref{sec:Simulations} to examine the effect of stratification and $Pr_m$ on the SSD onset criterion. We feel it helpful to introduce this result early, as it naturally follows the previous phenomenological discussion and is understandable without detailed reference to the simulation setup.

\begin{figure}
    \centering
    \includegraphics[width=\linewidth]{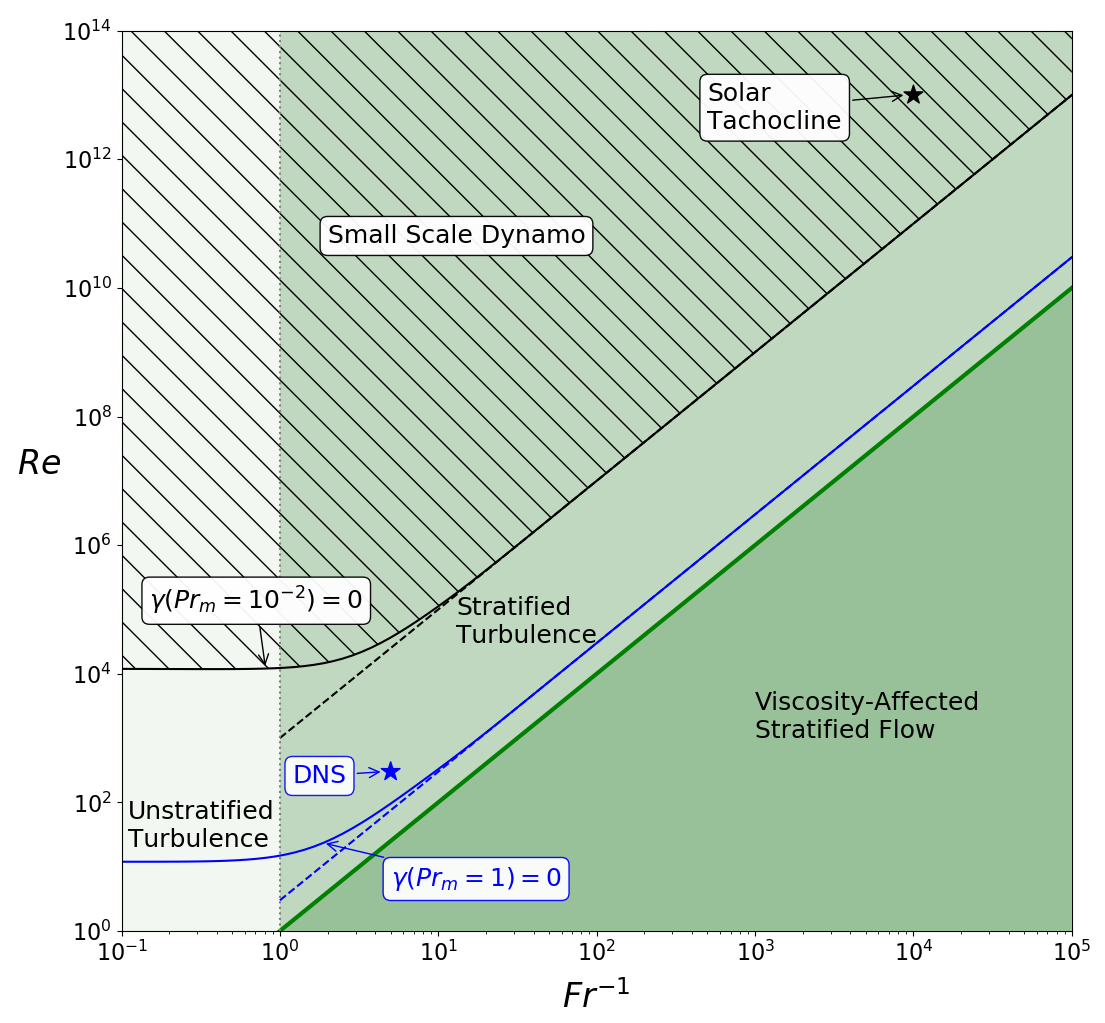}
    \caption{Small-scale dynamo instability diagram extended to stellar values of $Re$, $Fr$, and $Pr_m$ for $Pr=1$ based on the interpretation that $Rb_m^c$ is the correct onset criterion for $Pr_m<1$. The solid green line follows the $Rb=1$ ($k_O=k_\nu$) scaling, and regions with different shades of green mark different turbulence regimes. The black hashed region marks where the dynamo is unstable ($\gamma>0$), bounded by the solid black curve of the dynamo onset boundary for the representative solar tachocline value of $Pr_m=10^{-2}$. The solid blue curve marks the onset boundary for $Pr_m=1$ extended from the DNS. The dashed black and blue lines follow the asymptotes $Rb_m=Rb_m^c=9$ and $Rb_m=Rb_m^c=3$ for the $Pr_m=10^{-2}$ and $Pr_m=1$ cases, respectively.}
    \label{fig:ReFrContourSketch}
\end{figure}
\subsection{Stably Stratified SSD Onset Criterion}\label{sec:OnsetCriterion}
The onset criterion at a fixed $Pr_m$ can be defined as the critical Reynolds number $Re^c(Fr)$ that satisfies $\gamma(Re^c,Fr)=0$, where $\gamma$ is the SSD growth rate. In other words, any larger Reynolds number $Re>Re^c$ at constant stratification $Fr$ will lead to instability $\gamma>0$. Determining the onset criterion requires an expensive 2D scan of $Re-Fr$ space for each $Pr_m$ in order to reveal the scaling relationship when the dynamo turns on as stratification is decreased. For example, in the $Pr_m\geq 1$ case, a scaling
\begin{equation}
    Re^c\sim Fr^{-4/3},
\end{equation} 
implies $k_\nu\sim k_b$, while 
\begin{equation}
    Re^c\sim Fr^{-2},
\end{equation} 
implies $k_\nu\sim k_O$. An intermediate scaling would satisfy
\begin{equation}\label{eq:ThreshBoundaryPr1}
    Re^c\sim Fr^{-m} \;\;\; (4/3<m<2).
\end{equation} 
In Section \ref{sec:Simulations} we show that $m=2$ is the best fit for $Pr_m=1$ and $Pr_m=8$, implying that $k_O\sim k_\nu$ at onset or, in other words, that there is likely a critical buoyancy Reynolds number $Rb^c$ for $Pr_m\geq1$. For lower $Pr_m$, computational resources limit a full scan of $Re-Fr$ space, but a scan across a single value of $Re$ at $Pr_m=0.25$ shows that the dynamo shuts off when $k_O<k_\eta$. This suggests that $k_O\sim k_\eta$, as opposed to $k_b\sim k_\eta$, controls the dynamo onset at low $Pr_m$, which implies a critical magnetic buoyancy number $Rb_m^c>1$ for $Pr_m\leq1$. A more detailed analysis is shown in Section \ref{sec:Results}.

We extrapolate the scalings suggested by simulation results to stellar parameters in the sketch of the $Re-Fr$ plane shown in Figure \ref{fig:ReFrContourSketch}. The three turbulence regimes are colored with shades of green and superimposed with $Pr_m=1$ (blue) and $Pr_m=10^{-2}$ (black) dynamo onset curves (labeled $\gamma=0$). The SSD instability regions ($\gamma>0$) lie above the $\gamma=0$ curves and, for clarity, only the $Pr_m=10^{-2}$ instability region is marked by the hashed black lines. The main effect of lowering the $Pr_m$ is to raise the dynamo onset curves, whose horizontal portion for $Fr^{-1}\lesssim1$ is set by the y-intercept $Re^c=Pr_m^{-1}Rm^{c}$ and whose asymptotic portion for $Fr^{-1}\gg1$ is set by $Re^c=Pr_m^{-1}Fr^{-2}Rb_m^{c}$. When examining conditions in the Sun in Section \ref{sec:ApplicationToTC}, we find that the $(Re,Fr)$ values in the solar tachocline are plausibly inside the $Pr_m=10^{-2}$ SSD instability region. Note that the extrapolation to the astrophysical parameter regime assumes that the scaling found in the computationally accessible regime is asymptotic, which appears to be case for the values of $Pr_m=1,8$ detailed in Section \ref{sec:Simulations}.

\begin{figure}
    \centering
    \includegraphics[width=\linewidth]{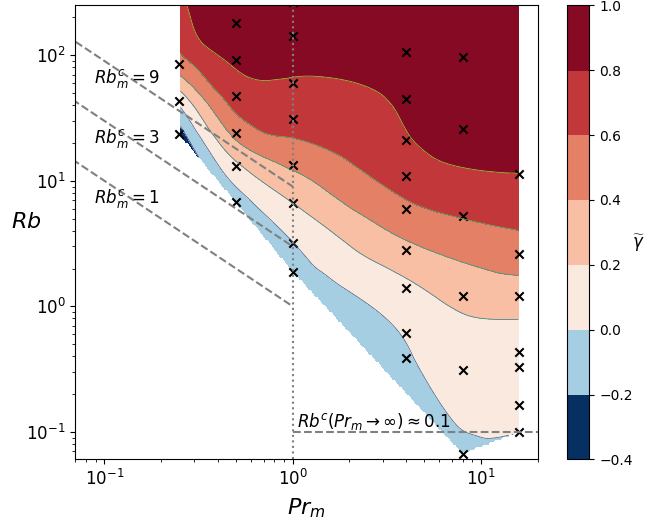}
    \caption{Contour plot of normalized growth rate $\widetilde \gamma$ in the $Rb-Pr_m$ plane using simulation sets $1$, $2$, $8$, $11$, $16$, and $18$ from Table \ref{tab:SimParam}. The SSD onset criterion curve $Rb^c(Pr_m)$ is seen as the boundary between white and blue contours. Dashed lines mark potential asymptotes of $Rb^c(Pr_m)$, which scale with $Rb$ at high $Pr_m$ and with $Rb_m$ at low $Pr_m$. The vertical dotted line marks the $Pr_m=1$ separation. }
    \label{fig:Rbcrit}
\end{figure}

When considering the stratified SSD onset criterion in the space of $Re-Fr-Pr_m$, the results suggest that all the relevant information can be represented in the $Rb-Pr_m$ plane instead of separate $Re-Fr$ planes at each $Pr_m$. The stratified SSD onset criterion is then determined by the curve $Rb^c(Pr_m)$. Combining sets of simulations varying $Fr$ at fixed $Re$ across the computationally accessible values of $0.25\leq Pr_m\leq16$, we generate a contour plot in the $Rb-Pr_m$ plane shown in Figure \ref{fig:Rbcrit} of the normalized growth rate $\widetilde{\gamma}=\gamma/\gamma_0$, where $\gamma_0$ is the unstratified growth rate with all other parameters fixed. The boundary between the light blue and white contours reveals the SSD onset criterion curve $Rb^c$ as a function of $Pr_m$. Dashed lines mark potential asymptotes in the low/high $Pr_m$ limits since one might expect $Rb^c$ and $Rb_m^c$ to become independent of $Pr_m$ (analogously to $Rm^c$) for $Pr_m\gg1$ and $Pr_m\ll1$, respectively. At higher $Pr_m>4$, the onset curve begins to flatten and suggests $Rb^c(Pr_m\rightarrow \infty)\simeq 0.1$; however, it is difficult to be conclusive with only two values of $Pr_m$. At lower $Pr_m<1$, $Rb_m^c$ increases with decreasing $Pr_m$ up to $Rb_m^c\approx 9$ at $Pr_m=0.25$, which is expected since $Rm^c$ increases for $Pr_m<1$ \citep{Iskakov2007}. If the $Rb_m^c$ curve qualitatively follows the $Rm^c$ curve for $Pr_m<1$, it is possible that $Rb_m^c$ decreases and plateaus after $Pr_m\lesssim 0.1$, meaning $Rb_m^c\approx9$ could be near the upper bound for $Rb_m^c(Pr_m\rightarrow 0)$. Unfortunately it is not possible to simulate $Pr_m< 0.25$ or $Pr_m>16$ with available resources due to the difficulty of resolving the three scale separations between stratification, resistive, and viscous scales.

\begin{table*}
     \centering
     \begin{center}
    \begin{tabular}{|c|c|c|c|c|c|c|}
        \hline
         Set&$Pr_m$ &  $Re$&  $Fr^{-1}$ & $N_xN_yN_z$ & $\nu^{-1}$ & $N^2$  \\[0.5ex]
         \hline\hline
         \multirow{1}{*}{1} & \multirow{1}{*}{0.25} & \multirow{1}{4.5cm}{\centering\{343, 322, 322, 341, 361\}} & \multirow{1}{3.5cm}{\centering\{0.2, 1.0, 1.9, 2.8, 3.9\}} & \multirow{1}{*}{$448^3$}   & \multirow{1}{*}{16000}  & \multirow{1}{4cm}{\centering\{1, 16, 64, 128, 256\}} \\
         
         \multirow{2}{*}{2} & \multirow{2}{*}{0.5}  & \multirow{2}{4.5cm}{\centering\{176, 171, 175, 178, 186, 201, 179\}}  & \multirow{2}{3.5cm}{\{\centering0.2, 1.0, 1.4, 1.9, 2.8, 3.9, 5.1 \}}& \multirow{2}{*}{$256^3$}   & \multirow{2}{*}{8000}   & \multirow{2}{4cm}{\centering\{1, 16, 32, 64, 128, 256, 384\}} \\
         & & & & & & \\
         
         \multirow{1}{*}{3} & \multirow{1}{*}{1}    & \multirow{1}{4.5cm}{\centering\{13, 13, 15, 15\}}  & \multirow{1}{3.5cm}{\centering\{0.6, 1.2, 1.9, 2.7\}} & \multirow{1}{*}{$256^3$}   & \multirow{1}{*}{500}  & \multirow{1}{4cm}{\centering\{1, 4, 16, 32\}} \\
         
         \multirow{2}{*}{4} & \multirow{2}{*}{1}    & \multirow{2}{4.5cm}{\centering\{32, 30, 30, 30, 30, 30\}}  & \multirow{2}{3.5cm}{\centering\{0.3, 0.6, 1.2, 1.7, 2.6, 3.7\}} & \multirow{2}{*}{$256^3$}   & \multirow{2}{*}{1000}  & \multirow{2}{4cm}{\centering\{1, 4, 16, 32, 64, 128\}} \\
         & & & & & & \\
         \multirow{2}{*}{5} & \multirow{2}{*}{1}    & \multirow{2}{4.5cm}{\centering\{54, 53, 51, 52, 52, 53, 54\}}  & \multirow{2}{3.5cm}{\centering\{0.25, 0.5, 1.1, 1.6, 2.1, 3.2, 4.8\}} & \multirow{2}{*}{$256^3$}   & \multirow{2}{*}{2000}  & \multirow{2}{4cm}{\centering\{1, 4, 16, 32, 64, 128, 256\}} \\
         & & & & & & \\
         
         \multirow{2}{*}{6} & \multirow{2}{*}{1}    & \multirow{2}{4.5cm}{\centering\{92, 91, 92, 96, 98, 94, 99, 97, 93\}}  & \multirow{2}{3.5cm}{\centering\{0.2, 0.5, 1.0, 1.3, 2.0, 2.9, 3.9, 5.3, 6.3\}} & \multirow{2}{*}{$256^3$}   & \multirow{2}{*}{4000}  & \multirow{2}{4cm}{\centering\{1, 4, 16, 32, 64, 128, 256, 384, 512\}} \\
         & & & & & & \\
         
         \multirow{2}{*}{7} & \multirow{2}{*}{1}    & \multirow{2}{4.5cm}{\centering\{135, 132, 137, 133, 131, 138, 143, 131, 136\}} & \multirow{2}{3.5cm}{\centering\{0.2, 0.5, 0.9, 1.4, 2, 2.9, 4.1, 5.8, 8.3\}} & \multirow{2}{*}{$256^3$}  & \multirow{2}{*}{6000} & \multirow{2}{4cm}{\centering\{1, 4, 16, 32, 64, 128, 256, 512, 1024\}} \\
         & & & & & & \\
         
         \multirow{2}{*}{8} & \multirow{2}{*}{1}    & \multirow{2}{4.5cm}{\centering\{178, 170, 178, 179, 181, 177, 182, 169, 180\}} &\multirow{2}{3.5cm}{\centering\{0.2, 0.5, 0.9, 1.4, 2, 2.8, 4.1, 5.9, 8\}}& \multirow{2}{*}{$256^3$}  & \multirow{2}{*}{8000}  & \multirow{2}{4.5cm}{\centering\{1, 4, 16, 32, 64, 128, 256, 512, 1024\}} \\
         & & & & & & \\
         
         \multirow{2}{*}{9} & \multirow{2}{*}{1}    & \multirow{2}{4.5cm}{\centering\{228, 228, 220, 229, 228, 223, 225, 218, 212, 216\}} &\multirow{2}{3.5cm}{\centering\{0.2, 0.4, 0.9, 1.3, 2, 2.7, 4.1, 5.7, 8.1, 10.7\}}& \multirow{2}{*}{$448^3$}  & \multirow{2}{*}{1000}  & \multirow{2}{4cm}{\centering\{1, 4, 16, 32, 64, 128, 256, 512, 1024, 2048\}} \\
         & & & & & & \\
         
         \multirow{2}{*}{10} & \multirow{2}{*}{1} & \multirow{2}{4.5cm}{\centering\{342, 345, 338, 351, 330, 348, 344, 335, 338, 314, 302\}} &\multirow{2}{3.5cm}{\centering\{0.2, 0.4, 0.9, 1.3, 2, 2.7, 4, 5.6, 7.9, 9.7, 11\}} & \multirow{2}{*}{$504^3$}   &\multirow{2}{*}{16000}  &\multirow{2}{4cm}{\centering\{1, 4, 16, 32, 64, 128, 256, 512, 1024, 1536, 2048\}}  \\
         & & & & & & \\
         
         \multirow{2}{*}{11} & \multirow{2}{*}{4}    & \multirow{2}{4.5cm}{\centering\{94, 97, 90, 92, 98, 99, 95, 93, 90\}}  &\multirow{2}{3.5cm}{\centering\{0.2, 1.0, 1.4, 2.1, 3, 4.1, 5.8, 8.1, 12\}}& \multirow{2}{*}{$256^3$}   & \multirow{2}{*}{4000} & \multirow{2}{4cm}{\centering\{1, 16, 32, 64, 128, 256, 512, 1024, 2048\}} \\
          & & & & & & \\
         
         \multirow{1}{*}{12} & \multirow{1}{*}{8}    & \multirow{1}{4.5cm}{\centering\{4.5, 4.1, 4.1, 3.9\}}  & \multirow{1}{3.5cm}{\centering\{0.6, 2.7, 5.5, 12\}}& \multirow{1}{*}{$256^3$}   & \multirow{1}{*}{100} & \multirow{1}{4cm}{\centering\{1, 16, 64, 256\}} \\
         
         \multirow{1}{*}{13} & \multirow{1}{*}{8}    & \multirow{1}{4.5cm}{\centering\{18, 17, 17, 18, 18, 18\}}  &\multirow{1}{3.5cm}{\centering\{0.4, 1.5, 3, 6.1, 13, 19\}}& \multirow{1}{*}{$256^3$}   & \multirow{1}{*}{500} & \multirow{1}{4cm}{\centering\{1, 16, 64, 256, 1024, 2048\}} \\
          
         \multirow{1}{*}{14} & \multirow{1}{*}{8}    & \multirow{1}{4.5cm}{\centering\{31, 29, 32, 31, 30, 32\}}  &\multirow{1}{3.5cm}{0.3, 1.3, 2.5, 5.2, 11, 22\}}& \multirow{1}{*}{$256^3$}   & \multirow{1}{*}{1000} & \multirow{1}{4cm}{\centering\{1, 16, 64, 256, 1024, 4096\}} \\
         
         \multirow{2}{*}{15} & \multirow{2}{*}{8}    & \multirow{2}{4.5cm}{\centering\{53, 53, 52, 54, 53, 53, 53\}}  & \multirow{2}{3.5cm}{\centering\{0.3, 1., 2.2, 4.8, 9.3, 19, 28\}}& \multirow{2}{*}{$256^3$}   & \multirow{2}{*}{2000} & \multirow{2}{4cm}{\centering\{1, 16, 64, 256, 1024, 4096, 8192\}} \\
          & & & & & & \\
         \multirow{2}{*}{16} & \multirow{2}{*}{8}    & \multirow{2}{4.5cm}{\centering\{93, 92, 95, 97, 93, 87, 93\}} &\multirow{2}{3.5cm}{\centering\{0.2, 1.0, 1.9, 4.3, 8.8, 16, 37\}}& \multirow{2}{*}{$448^3$}   & \multirow{2}{*}{4000} &  \multirow{2}{4cm}{\centering\{1, 16, 64, 256, 1024, 4096, 16384\}} \\
          & & & & & & \\
         \multirow{2}{*}{17} & \multirow{2}{*}{8}    & \multirow{2}{4.5cm}{\centering\{133, 132, 131, 136, 131, 123, 124, 123\}}  &\multirow{2}{3.5cm}{\centering\{0.2, 1.0, 2.1, 4.4, 8.3, 16, 29, 47\}}& \multirow{2}{*}{$504^3$}   & \multirow{2}{*}{6000} & \multirow{2}{4cm}{\centering\{1, 16, 64, 256, 1024, 4096, 16384, 32768\}} \\
          & & & & & & \\
         \multirow{2}{*}{18} & \multirow{2}{*}{16}   & \multirow{2}{4.5cm}{\centering\{92, 97, 98, 95, 91, 91, 95, 96\}} &\multirow{2}{3.5cm}{\centering\{0.3, 2.9, 6.1, 8.9, 14, 17, 24, 31\}}& \multirow{2}{*}{$448^3$}  & \multirow{2}{*}{4000} & \multirow{2}{4cm}{\centering\{1, 128, 512, 1024, 3072, 4096, 8192, 16384\}} \\
          & & & & & & \\
         \hline
    \end{tabular}
    \end{center}
    \caption{Table of simulation parameters. Each set corresponds to a series of simulations where only the Brunt-V\"ais\"al\"a frequency $N$ is varied. The resolution is denoted by $N_x N_y N_z$. The magnetic resistivity and thermal diffusivity (not shown) are given by $\eta=Pr_m^{-1} \nu$ and $\kappa=\nu$ ($Pr=1$), respectively.}
    \label{tab:SimParam}
\end{table*}

\subsection{Role of Thermal Prandtl number $Pr$}\label{sec:ThermalPr}
All simulations and most discussions in this paper pertain to the $Pr\sim1$ regime. However, radiative zones typically have extremely low $Pr$ that are also much smaller than their magnetic Prandtl numbers, $Pr\ll Pr_m<1$. Increased thermal diffusion relative to viscous dissipation increases the thermal dissipation scale below which buoyancy effects become less important and fluid motions more isotropic. A lower $Pr$ thus should lead to a more active SSD. A simple estimate can be made for how small $Pr$ must be to alter the SSD onset criterion. Balancing thermal diffusion and eddy turnover time for $Pr\ll1$, the thermal dissipation scale sits at $k_\kappa=Pr^{3/4}k_\nu$ and will significantly change the stratified turbulence picture when $k_\kappa< k_O$. Equivalently, when
\begin{equation}\label{eq:LowPr}
    Pr< Rb^{-1},
\end{equation}
velocity scales smaller than $k>k_\kappa$ will be more isotropic than in the $Pr\sim1$ case, resulting in an increased SSD growth rate and an extended parameter space of unstable dynamos (e.g. unstable when $k_\kappa< k_\eta$ instead). 

A lower $Pr$ can also affect the SSD depending on the driving mechanism(s) of the background stratified turbulence, for example, by enhancing horizontal and vertical shear instabilities \citep{zahn1974rotational,prat2016shear,lignieres2019turbulence,cope2020dynamics,garaud2020horizontal}, damping internal gravity waves, and/or affecting the nature of nearby convection zones and associated convective overshoot dynamics \citep{Elliott_2000,miesch2005large,Brun_2011,featherstonePrEffect}. We leave studies of the effect of low $Pr$ on the SSD for future work.

\section{Simulations}\label{sec:Simulations}

\subsection{Setup}\label{sec:setup}
We use SNOOPY \citep{SNOOPY2005A&A}, a 3D pseudospectral code, with low-storage third-order Runge-Kutta time stepping and 3/2 de-aliasing to carry out DNS of the incompressible MHD Boussinesq equations:

\begin{equation}
\partial_t \textbf{u}+\textbf{u}\cdot\nabla \textbf{u}=-\nabla p-N^2\theta\hat{z}+\textbf{B}\cdot\nabla \textbf{B}+\nu\nabla^2\textbf{u}+\sigma_f,
\end{equation}

\begin{equation}
\partial_t \theta+\textbf{u}\cdot\nabla \theta=u_z+\kappa\nabla^2\theta,
\end{equation}

\begin{equation}
\partial_t \textbf{B}+\textbf{u}\cdot\nabla \textbf{B}=\textbf{B}\cdot\nabla\textbf{u}+\eta\nabla^2\textbf{B},
\end{equation}
\begin{equation}
\nabla \cdot \textbf{u}=0, \; \nabla \cdot \textbf{B}=0,
\end{equation}
where $\textbf{u}$ is the velocity field, $\theta$ is the buoyancy variable, $\textbf{B}$ is the magnetic field normalized by $\sqrt{4\pi \rho_0}$, $\rho_0$ is the constant plasma density, and $\sigma_f$ is the kinetic forcing term. Fluid velocities in stellar radiative zones are highly subsonic, implying that compressibility effects will not be important for the dynamo \citep{federrath2011mach} and justifying use of the Boussinesq approximation. 

All simulations use triply periodic, cubic boxes ($L=1$) and the Prandtl number $Pr=\frac{\nu}{\kappa}=1$ is kept fixed, while the remaining parameters $Pr_m$, $Re$, and $Fr$ are varied throughout the paper. We use isotropic, nonhelical, time-correlated forcing with wavenumbers $\frac{k}{2\pi}\in [2.25,3.75]$ and correlation time $\tau_c=0.3 \sim l_i/u_{\rm rms}$ ($u_{\rm rms}\sim1$ in all simulations). We have compared with forcing of only horizontal wavenumbers (not shown) as is often implemented in geophysical applications and have found little effect on turbulent spectra for wavenumbers $k>k_i$. The smallest scales are known to primarily contribute to the SSD growth rate, and as a result the SSD ends up being insensitive to the nature of the large-scale forcing, although SSD saturation and the large-scale dynamo will likely have a stronger dependence.  

\begin{figure*}
    \centering
    \includegraphics[width=\linewidth]{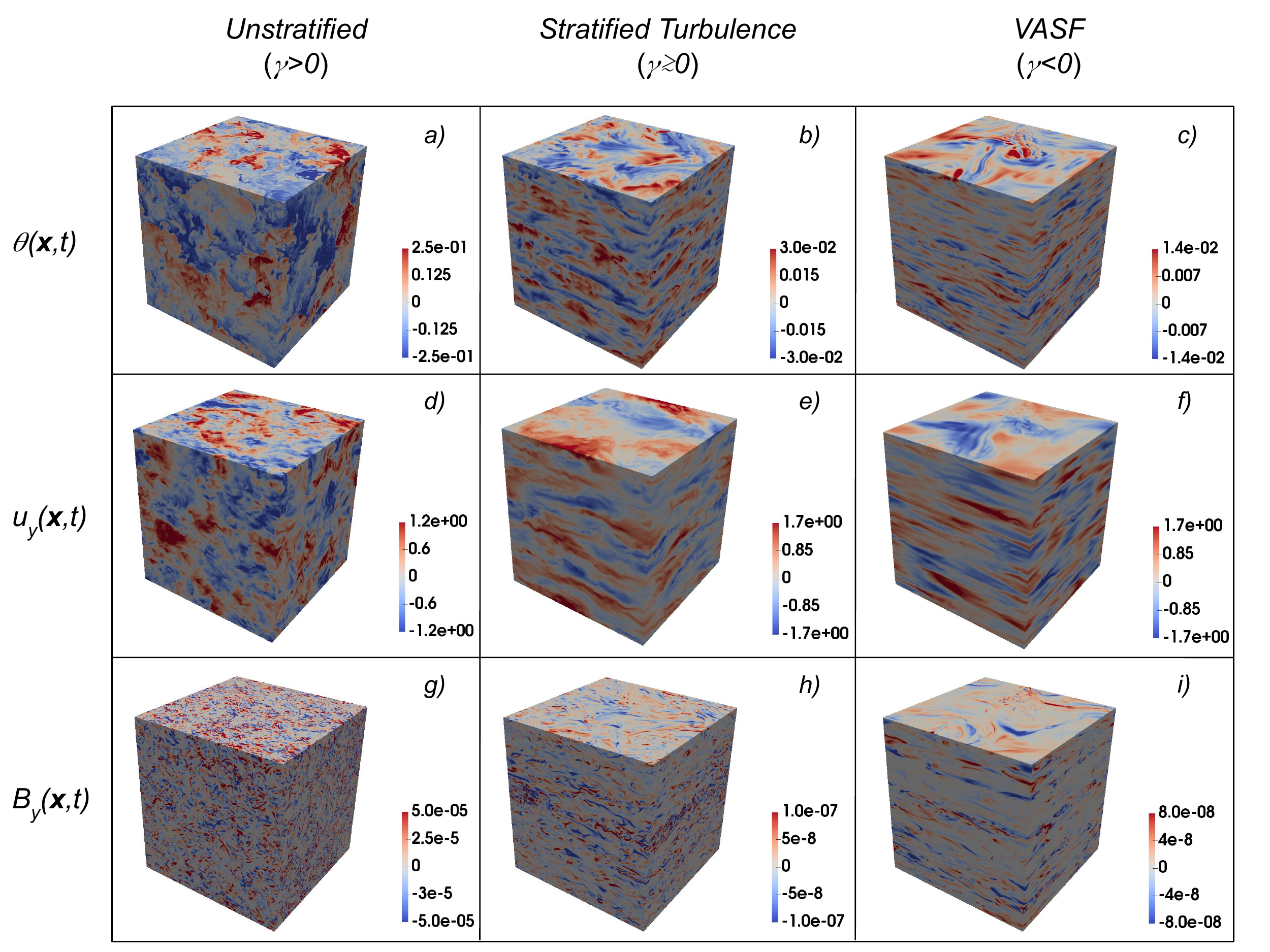}
    \caption{Snapshots of physical space surface plots at $t= 12 \tau_c$ for representative simulations at $Re\approx220$ and $Pr_m=Pr=1$. Top, middle, and bottom rows correspond to the buoyancy, horizontal velocity, and horizontal magnetic fields, respectively. Left, center, and right columns correspond to the three turbulence regimes with Froude numbers $Fr^{-1}=0$, $Fr^{-1}\approx8$, and $Fr^{-1}\approx16$, respectively.}
    \label{fig:Cubes}
\end{figure*}
Table \ref{tab:SimParam} shows the parameters used for all sets of simulations presented in the paper. In a set of simulations, the Brunt-V\"ais\"al\"a frequency $N^2$ is varied while the amplitude of the forcing term $\sigma_f$ is adjusted to keep $u_{\rm rms}\approx 1$.

At the beginning of the simulations, isotropic forcing quickly excites all wavenumbers, and the stratified background turbulence reaches a steady state within time $t\approx \!5\tau_c$. Simulations are integrated in time until the magnetic energy either has grown from an initialized weak field ($|B|\approx \!10^{-8}$) by several orders of magnitude or $t\approx 30\tau_c$. The magnetic energy always stays below the energies of the viscous eddies, keeping the simulation in the kinematic dynamo regime. Growth rates are then calculated from a linear fit of $\log\left(E_B(t)\right)$ vs $t$ for $t>5\tau_c$, where $E_B$ is the total magnetic energy.

\subsection{Diagnostics}\label{sec:Diagnostics}
\subsubsection{Anisotropy Diagnostics}
Several spectral diagnostics are implemented to characterize the departure from isotropy of both the components and the angular energy spectra of the velocity and magnetic fields. Here we write out the diagnostics for the velocity field $\textbf{u}(\textbf{x})$ whose Fourier transform is denoted as $\hat{\textbf{u}}(\textbf{k})$.

To study the component anisotropy, the energy spectra are simply split into the contribution from each component:

\begin{equation}
    E^i_u(k)=\frac{1}{2}\sum_{|\mathbf{k}|\in [k-\pi,k+\pi]}|\hat{u}_i(\mathbf{k})|^2,
\end{equation}
with total energy given by $E_u(k)=\sum_i E^i_u(k)$ (similarly for magnetic energy components $E_B^i(k)$).  

The remaining anisotropy can manifest as a variation of the energy spectra in angular spectra with respect to the angle $\theta=\sin^{-1}(k_z/k)$. To study the angular dependence of the energy spectra, following \cite{Lang2019}, we bin each spherical angular spectra of the energy spectra further into 2M latitudinal bands with equal angular spacing $\Delta \theta=\pi/2M$. Denote $O_{k,i}$ as the set of wavenumbers with $|\textbf{k}|\in[ k-\pi,k+\pi ]$ and angle $\theta\in\pm[\theta_i,\theta_{i+1}]$ measured from the horizontal plane with $1\leq i\leq M$. The $i$th angular energy spectrum is then given by
        
\begin{equation}
    E_u(k,i)=\frac{1}{m_i}\sum_{\mathbf{k}\in O_{k,i}} \frac{1}{2}\hat{u}_j(\mathbf{k})\hat{u}_j^*(\mathbf{k}),
\end{equation}
with weights $m_i=M |O_{k,i}|/\sum_{j=1}^M|O_{k,j}|$ to ensure that all of the angular spectra are equal in the isotropic limit (similarly for the magnetic spectra $E_B(k,i)$). 

A $k$-dependent dimensionless measure of the angular spectra anisotropy can then be given by 
\begin{equation}
a_u(k)=\sigma_u(k)/\mu_u(k),
\end{equation}
the standard deviation $\sigma_u^2(k)=M^{-1}\sum_i(E_u(k,i)-\mu_u(k))^2$ divided by the mean $\mu_u(k)=M^{-1}\sum_iE_u(k,i)$ of the angular bins (similarly for the magnetic field $a_B(k)$). Purely isotropic turbulence would have $a_u(k)\approx0$.
\subsubsection{Dimensionless Parameters and Scales}
The growth rate is studied with respect to the relative quantitative separation of stratification scales to dissipation scales, which require a concrete measure of the dimensionless parameters ($Re,Fr$) from the simulation output. Because the ratio of thermal to viscous dissipation $\epsilon_p/\epsilon_k$ at fixed energy input $\epsilon=\epsilon_k+\epsilon_p$ varies with $Fr$, the viscous scale $k_\nu=(\epsilon_k/\nu^3)^{1/4}$ would vary at fixed $Re$ if the Reynolds number was defined as usual relative to the unchanging, large-scale parameters $Re^{\rm std}=u_{\rm rms}/\nu k_i$ \citep{pouquet_2018}. This would break the scaling $k_\nu\sim Re^{3/4}k_i$ that is important for our analysis. Instead, we define the Reynolds number 
\begin{equation}
Re=\epsilon_k^{1/3}l_i^{4/3}/(2\pi \nu),    
\end{equation}
based on the viscous energy dissipation rate, which gives the standard definition of $Re$ when $\epsilon_p=0$. $\epsilon_k$ is calculated from the simulation

\begin{equation}
    \epsilon_k=\nu\int|\nabla\times\textbf{u}(\textbf{x})|^2 d^3x,
\end{equation}
allowing for a direct measure of the effective Reynolds number.  We find $Re^{\rm std}/Re$ increases systematically with $Fr^{-1}$, as expected because at lower $Fr$ a larger fraction of the total energy injection is directed towards the buoyancy cascade. However, across all of our simulations, $Re^{\rm std}/Re$ is never larger than $\simeq2$ or smaller than $\simeq1.4$, and our conclusions would be mostly unchanged if we used $Re^{\rm std}$ rather than $Re$ in our scaling analyses.

Stratification scales are calculated using $k_b/2\pi=N/u_{\rm rms}$ and  $k_O/2\pi=(N^3l_i/u^3_{\rm rms})^{\frac{1}{2}}$ where $N$ is a simulation input and $u_{\rm rms}$ and $l_i$ are measured from the simulation as
\begin{equation}
    u_{\rm rms}=\left[2\int_0^{\infty} E(k)dk\right]^{1/2},
\end{equation}
\begin{equation}
    l_i=2\pi\frac{\int_0^{\infty} k^{-1}E(k)dk}{\int_0^{\infty} E(k)dk}.
\end{equation}
These definitions are standard and allow for the exact scaling relations $k_b=Fr^{-1}k_i$ and $k_O=Fr^{-3/2}k_i$. 
\subsubsection{Transfer Function Diagnostics}
\begin{figure*}
\centering
\includegraphics[width=0.75\linewidth]{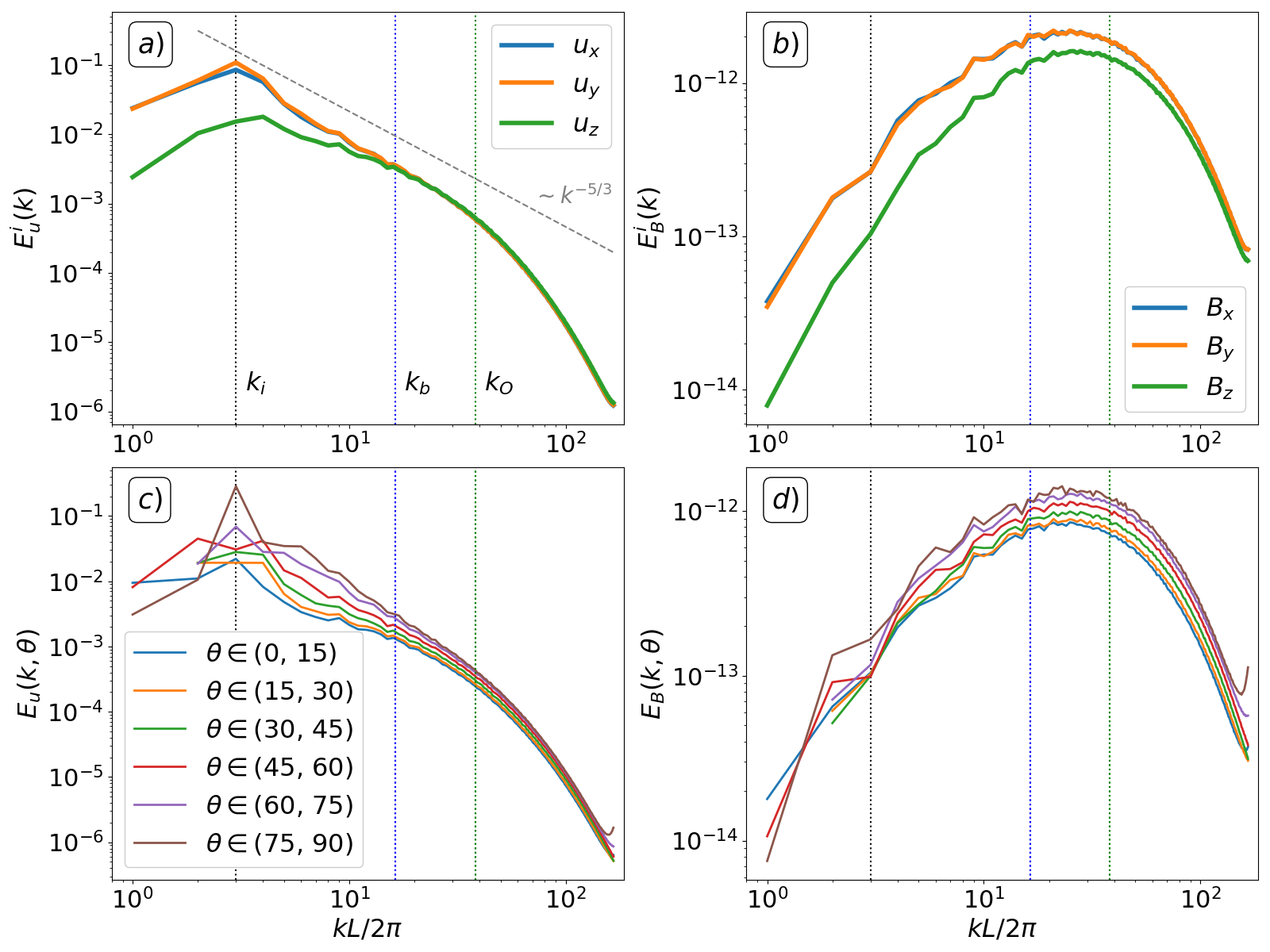}
\caption{Simulation with moderate stratification at $Re\approx 220$, $Fr^{-1}\approx4$, and $Pr_m=1$. Panels a) and b)  are spectra of individual components of the velocity and magnetic field, respectively. Panels c) and d) are angular energy spectra ($\theta=|\sin(k_z/k)^{-1}|$) of the velocity and magnetic field, respectively. See Section \ref{sec:Diagnostics} for details. 
\label{fig:SingleRunPr1}}
\end{figure*}

\begin{figure}
\centering
\includegraphics[width=\linewidth]{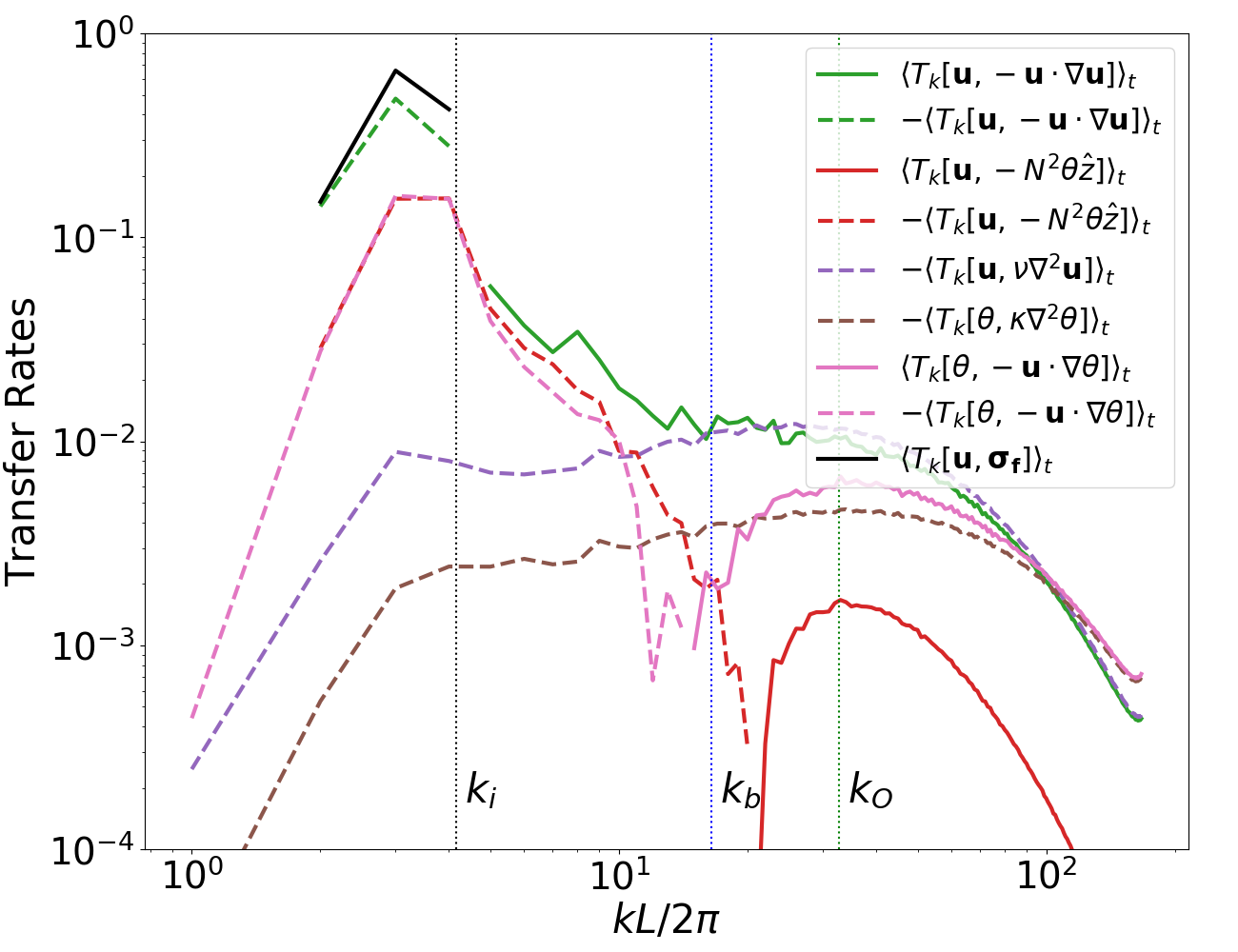}
\caption{Net energy transfer rates for terms in the momentum and buoyancy equations for the representative $Pr_m=1$, $Re\approx 220$, $Fr^{-1}\approx4$ simulation. Averages $\langle T_k[\textbf{V},\textbf{A}]\rangle_t$ are taken in time over a time interval $3\tau_c$ at the end of the simulation. Solid lines correspond to net energy flow into shell $k$ and dashed lines correspond to net energy flow out of shell k. The integral, buoyancy, and Ozmidov wavenumbers are marked by the vertical dashed black, blue, and green lines, respectively.}\label{fig:TransferKtoK}
\end{figure}

\begin{figure*}
\centering
\includegraphics[width=0.75\linewidth]{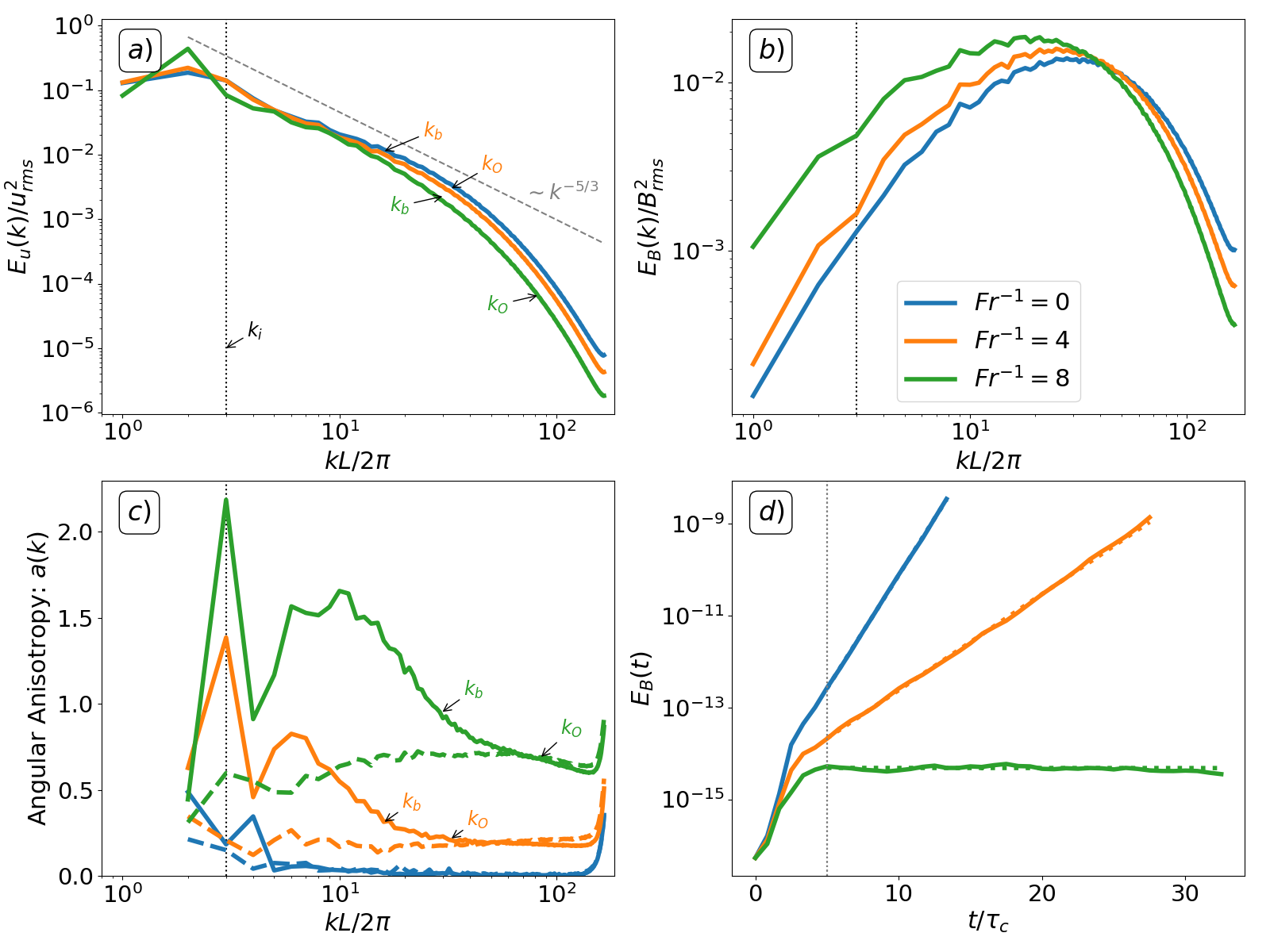}
\caption{Comparison of weakly, moderately, and strongly stratified simulations at $Re\approx 220$ and $Pr_m=1$. a) Normalized total energy spectra. b) Normalized magnetic energy spectra. c) Dimensionless angular energy spectra anisotropy (see Section \ref{sec:Diagnostics}). d) Magnetic energy versus time. Vertical gray line marks steady state turbulence after which the growth rate is calculated and fit, shown with dashed lines of the same color. 
\label{fig:ComparisonPr1}}
\end{figure*}
\begin{figure}
\includegraphics[width=\linewidth]{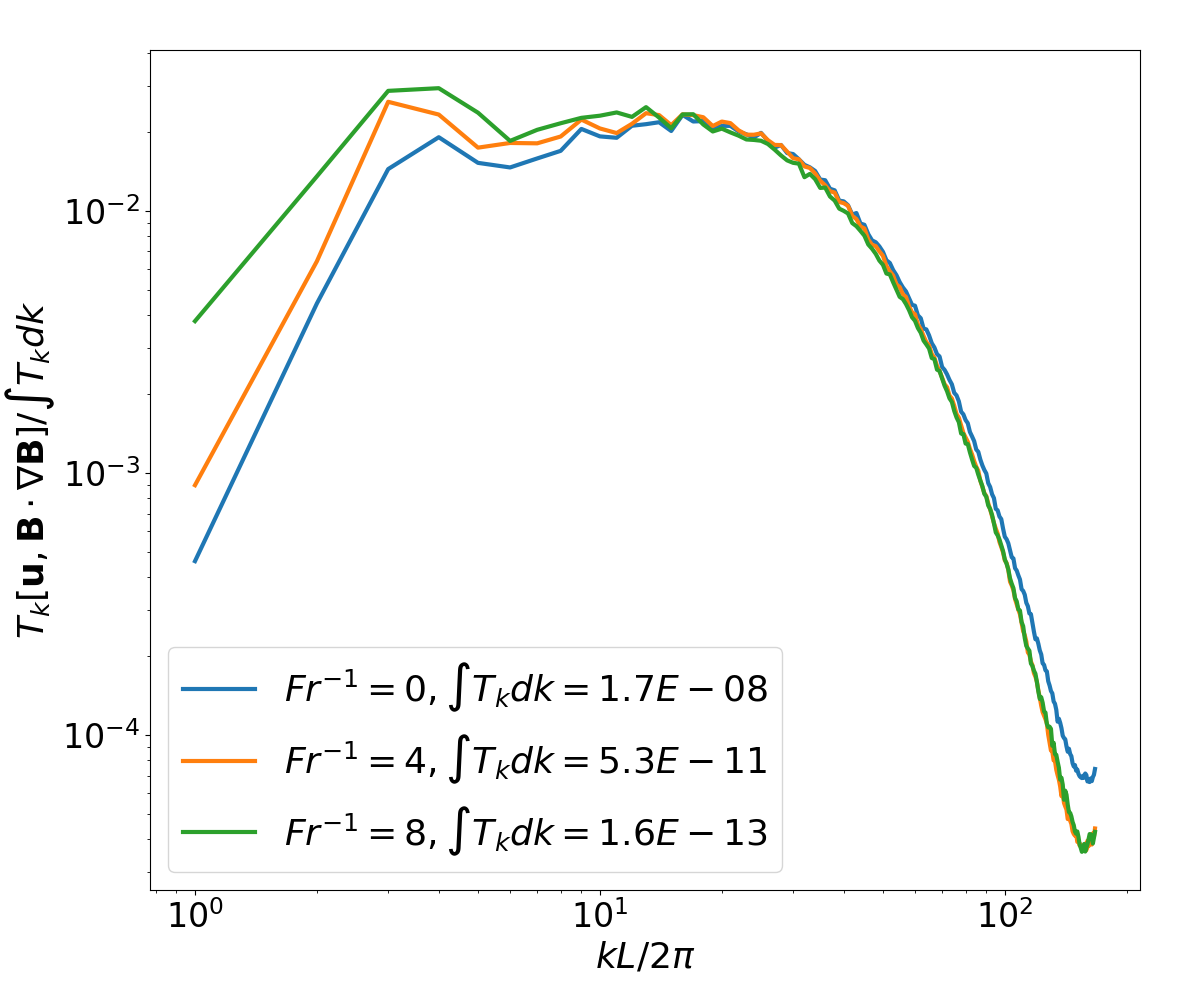}
\caption{Energy-transfer rate from velocity shell $k$ into magnetic energy normalized by the total kinetic to magnetic energy-transfer rate for the $Re\approx 220$, $Pr_m=1$ simulations. }\label{fig:TransferKtoM}
\end{figure}

Following \cite{alexakis2005shell,beresnyak2012universal,grete2017energy,St-Onge2020}, we implement shell-filtered energy transfer functions $T_k[\mathbf{V},\textbf{A}]$,
\begin{equation}
    T_k[\mathbf{V},\textbf{A}]=\sum_{q\in O_k} \mathbf{V_q}\cdot \mathbf{A_q},
\end{equation}
where $O_k$ is the set of wavenumbers $|\textbf{k}|\in[ k-\pi,k+\pi ]$, to examine the scale-by-scale energy balance of terms in the momentum, induction, and buoyancy equations corresponding to choices of $\mathbf{V}=\mathbf{u}$, $\mathbf{V}=\mathbf{B}$, and $\mathbf{V}=N^2\theta$, respectively. $T_k[\mathbf{V},\textbf{A}]$ measures the net rate of energy transfer into Fourier shell $k$ of $\mathbf{V}$ due to the term $\textbf{A}$ in the corresponding equation. For example, $-T_k[\mathbf{u},-N^2\theta \hat{z}]$ and $-T_k[\mathbf{u},\textbf{B}\cdot\nabla \textbf{B}]$ measure the rate of conversion of kinetic energy in velocity Fourier shell $k$ into buoyancy potential energy and magnetic energy, respectively. 

\subsection{Results}\label{sec:Results}
Physical space plots of representative simulations in the three turbulence regimes are shown in Figure \ref{fig:Cubes}. The stratified turbulence regime clearly has vertical layering with intermittent bursts of turbulence where the magnetic field is primarily amplified. This is unlike the unstratified case, where the magnetic field appears uniformly spread out. In the VASF regime, the flow appears smooth, and the decaying magnetic field has a similar structure. In this section, we quantitatively study these patterns in detail for $Pr_m=1$, high $Pr_m$, and low $Pr_m$ values.

\subsubsection{$Pr_m=1$ Regime Results}\label{sec:Prm1}

Fixing $Pr_m=1$, we explore a numerically accessible range of the remaining 2D space to study the nature of the anisotropy in the magnetic spectrum and the behavior of the dynamo growth rate: 
\begin{equation}
    \gamma=\gamma(Re,Fr,Pr_m=1).
\end{equation}

\paragraph{Analyzing an Individual Simulation} Spectral diagnostics of a representative simulation at moderate stratification with $Re\approx220$ and $Fr^{-1}\approx4$ are shown in Figure \ref{fig:SingleRunPr1}. The forcing (integral), buoyancy, and Ozmidov scales are shown as black, blue, and green vertical dashed lines. Examining the fluid spectra component-wise, the energy in the $u_z(k)$ component (Figure \ref{fig:SingleRunPr1}a) strictly above the buoyancy scale $k\leq k_b$ is notably smaller than in the $u_x(k)$ and $u_y(k)$ components, cleanly demonstrating the suppression of vertical motions by stratification. The energy in the $B_z(k)$ (Figure \ref{fig:SingleRunPr1}b) component is likewise significantly lower than in the horizontal magnetic components, becoming more equipartitioned at smaller scales; however, the magnetic component anisotropy is robustly present below the buoyancy scale $k\geq k_b$ unlike in the fluid component spectra.

For the angular energy spectra, vertical wavenumbers ($k\approx k_z$) in the velocity field (Figure \ref{fig:SingleRunPr1}c) dominate in energy at large-scales, with the buoyancy wavenumber marking the transition where the angular spectra anisotropy visibly begins to decrease. The velocity-field angular spectra anisotropy drives a preferential growth of vertical wavenumber modes in the magnetic field (Figure \ref{fig:SingleRunPr1}d). The magnetic-field angular spectra anisotropy appears to be roughly constant across all scales even though the velocity field notably became more isotropic below the buoyancy scale. The angular spectra anisotropy is similar for all components individually (not shown).

We plot kinetic and buoyancy energy-transfer functions in Figure \ref{fig:TransferKtoK} to help understand the the flow of energy and reaffirm the role of the stratification scales. The forcing term $T_k[\textbf{u},\mathbf{\sigma_f}]$ supplies energy at the largest scales $k\approx k_i$ followed by a cascade down to smaller scales. At large and intermediate scales, $k\lesssim k_b$, the nonlinear (NL) kinetic energy transfer $T_k[\textbf{u},\textbf{u}\cdot\nabla \textbf{u}]$ is primarily channeled into buoyancy energy through $T_k[\textbf{u},-N^2\theta \hat{z}]$, which cascades to smaller buoyancy scales through the NL buoyancy advection term $T_k[\theta,-\textbf{u}\cdot\nabla \theta]$. A transition occurs in between $k_b<k<k_O$ where the dominant energy exchange switches to NL kinetic energy transfer balancing viscous dissipation $T_k[\textbf{u},\nu\nabla^2 \textbf{u}]$ for the momentum equation and NL buoyancy energy transfer balancing thermal dissipation $T_k[\theta,\kappa\nabla^2\theta]$ for the buoyancy equation. Note that a balance between the transfer functions of two terms does not mean their influence on the flow is of similar importance: a small transfer function can signal a net balance between the energy going into and out of the $k$ shell in question, even if the term has a strong effect on the flow (as occurs for e.g., $\textbf{u}\cdot\nabla \textbf{u}$). Overall, the balance between inertia and buoyancy for $k\lesssim k_b$ and inertia and viscosity for $k\gtrsim k_b$ in the transfer rates neatly aligns with the observed anisotropy for $k\lesssim k_b$ and quasi-isotropy for $k\gtrsim k_b$ in the anisotropy diagnostics.

\paragraph{Comparison of Simulations} At the same approximately fixed $Re\approx220$, we additionally compare with an unstratified ($Fr^{-1}=0$) and a stronger ($Fr^{-1}\approx 8$) stratification case in Figure \ref{fig:ComparisonPr1}. The velocity angular spectra anisotropy $a_u(k)$ (solid lines in Figure \ref{fig:ComparisonPr1}c) dramatically rises with increasing stratification. The general shape of $a_u(k)$ for $k>k_i$ takes on a peak followed by a steep decrease, in the middle of which sits the buoyancy wavenumber, confirming the observation in the $Fr^{-1}=4$ case in Figure \ref{fig:SingleRunPr1}c. The Ozmidov scale appears to roughly mark the scale at which the steep slope transitions into a shallower slope. Note that $a_u(k)$ is finite for $k\rightarrow k_\nu$ even at moderate stratification, for example, $a_u(kL/2\pi=100)\simeq 0.2$ for $Fr^{-1}=4$, qualitatively agreeing with hydrodynamic simulations of \cite{Lang2019}. 

Due to the increasing anisotropy at the viscous scales, the dynamo growth rate drops sharply with increasing stratification, as shown in Figure \ref{fig:ComparisonPr1}d. The angular spectra anisotropy $a_B(k)$ of the dynamo-generated magnetic field (dashed lines in Figure \ref{fig:ComparisonPr1}c) likewise increases with stratification but behaves differently than $a_u(k)$. For a given $Fr$, $a_B(k)$ stays roughly constant across all $k$ at the same value as $a_u(k)$ near the viscous scales. This suggests that the anisotropy of the generated magnetic field is dominantly controlled by the most viscous eddies in the kinematic regime, as expected since the viscous scales dominate SSD growth (see Section \ref{sec:IntroSSD}).

Note that the normalized total magnetic spectrum (Figure \ref{fig:ComparisonPr1}b) to shifts slightly toward lower $k$. This is because the tail of the total velocity energy spectrum in Figure \ref{fig:ComparisonPr1}a moves slightly toward lower $k$ (i.e. the viscous scale increases) with increasing stratification at fixed $u_{\rm rms}\approx1$ and is the reason we have chosen to base the Reynolds number $Re$ on the exact value of $\epsilon_k$ as described in Section \ref{sec:Diagnostics}.
 
The normalized transfer of kinetic energy in velocity shell $k$ into magnetic energy is shown separately in Figure \ref{fig:TransferKtoM} for the three stratifications $Fr^{-1}=0,4,8$. The curves have little variation with stratification except for a slight relative increase at the large-scales. The main change with stratification is that the total energy transfer rate $-\int T_k[\textbf{u},\textbf{B}\cdot\nabla \textbf{B}]$ sharply decreases. In combination with the approximate scale independence of both the component-wise anisotropy and $a_B(k)$ shown in Section \ref{sec:Prm1}, the self-similarity of the transfer curve further suggests that the growth rate is primarily set by the velocity field at a particular scale and the sharp decrease in total energy transfer is due to the increased anisotropy at that scale. A more thorough analysis of the kinetic to magnetic transfer rates (e.g. shell to shell, component-wise) is left for future study. 

\begin{figure}[h]
\includegraphics[width=\linewidth]{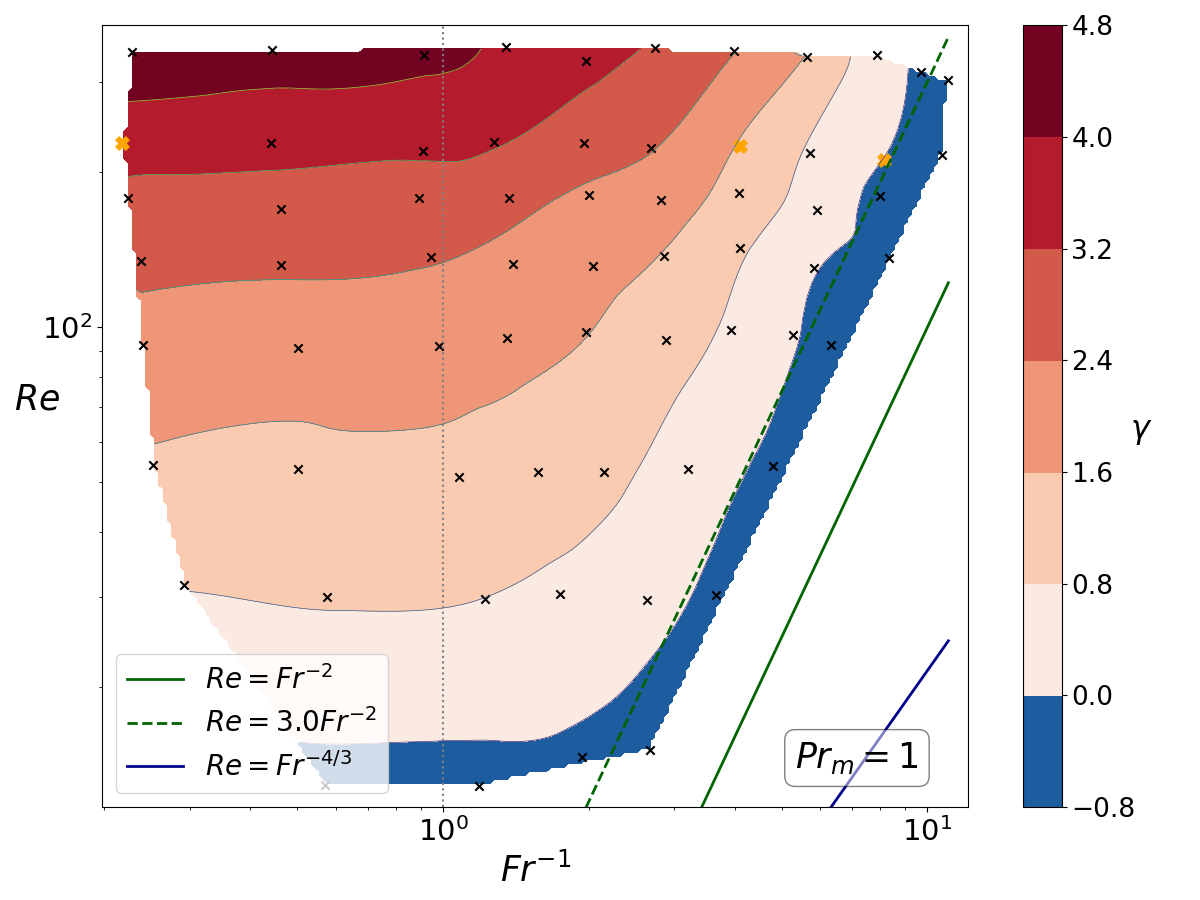}
\caption{ Contour plot of the dynamo growth rate $\gamma$ in the space of $Re$ vs $Fr^{-1}$ for $Pr_m=1$ using sets $3-10$ from Table \ref{tab:SimParam}. The blue and green lines are the scalings $Re^{3/4}=Fr^{-1}$ and $Re^{3/4}=Fr^{-3/2}$ corresponding to the wavenumber scalings $k_\nu=k_b$ and $k_\nu=k_O$, respectively. Black crosses are individual simulations, and bold orange crosses correspond to simulations analyzed in Figure \ref{fig:SingleRunPr1} and \ref{fig:ComparisonPr1}. Note that the onset curve at low $Re$ is horizontal, corresponding to the critical $Re^{c}\approx 20$ needed to excite the unstratified $Pr_m=1$ dynamo.  \label{fig:ContourPlotPrm1}}
\end{figure}

\paragraph{The Dynamo Onset Criterion} Next, we would like to understand the dynamo onset curve $Re^{c}(Fr)$ that satisfies $\gamma(Re^{c},Fr)=0$. The asymptotic slope of the onset boundary at higher $Re$ determines if $k_\nu=k_\eta$ scales with either $k_b$, $k_O$, or an intermediate scale when the dynamo shuts off. This leads to the scaling relation Equation \eqref{eq:ThreshBoundaryPr1} $Re^c\sim Fr^{-m}$ discussed in Section \ref{sec:OnsetCriterion}. We show the contour plot of $\gamma(Re,Fr)$ in Figure \ref{fig:ContourPlotPrm1} with the $\gamma=0$ curve as the boundary between the white and blue regions. The contour plot is generated by decreasing $Fr$ at roughly fixed values of $Re$ until the growth rate turns negative, revealing the dynamo onset boundary.

\begin{figure}[h]
\includegraphics[width=\linewidth]{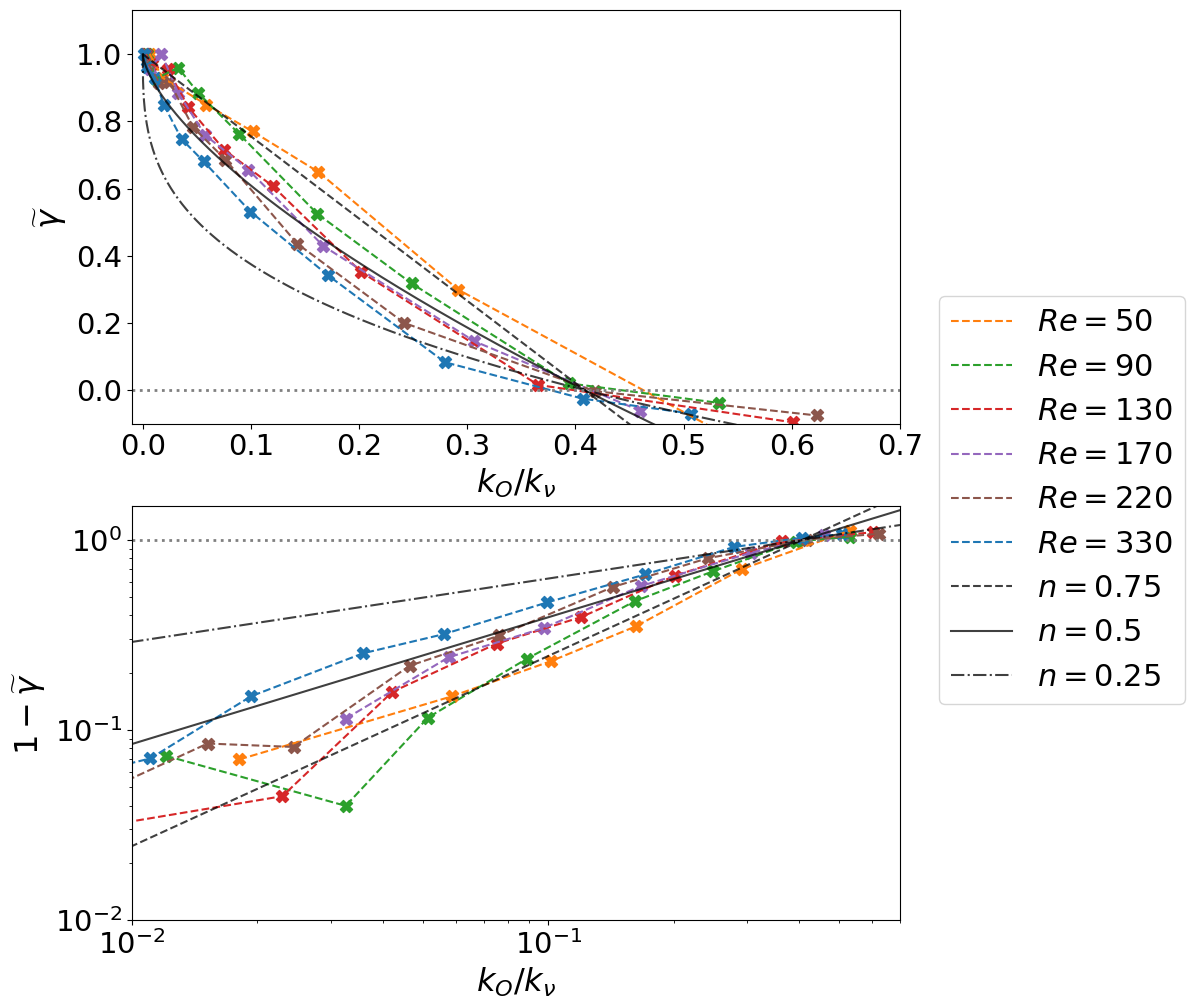}
\caption{Normalized growth rate $\widetilde{\gamma}$ at $Pr_m=1$ for different $Re$ versus the scale separation between Ozmidov and viscous scales $k_O/k_\nu=Rb^{-3/4}$. Top: linear-linear plot of $\widetilde{\gamma}$ versus $k_O/k_\nu$. Bottom: log-log plot of $1-\widetilde{\gamma}$ versus $k_O/k_\nu$ in order to look for potential scaling near criticality. Black curves correspond to empirical fits of Equation \eqref{eq:EmpiricalFitPrm1}. }
\label{fig:Fig4_NormalizedGammaPr1}
\end{figure}

\begin{figure*}
\centering
\includegraphics[width=0.75\linewidth]{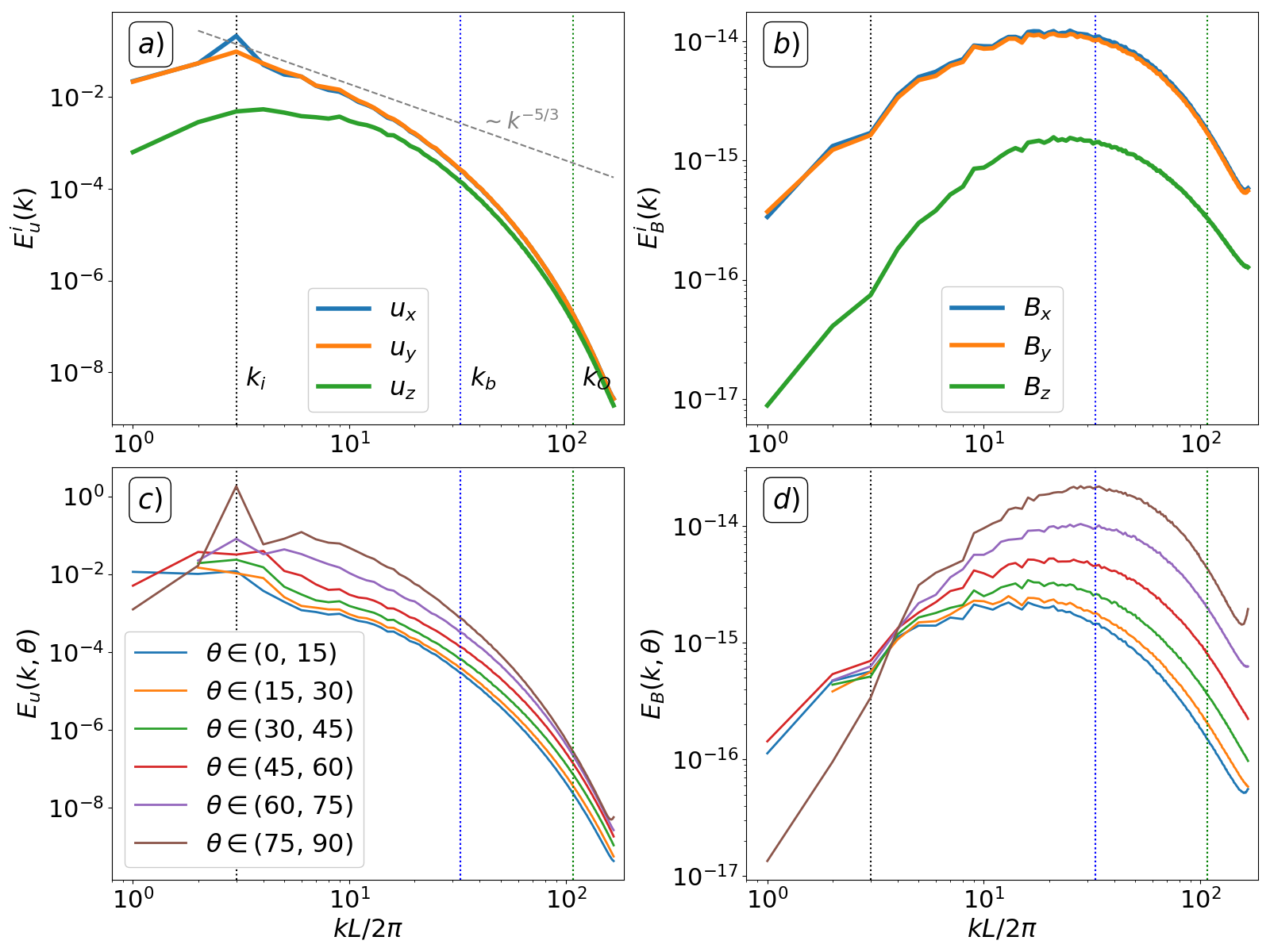}
\caption{ Spectral diagnostics of the $Pr_m=8$ simulation with strong stratification, $Re\approx 90$, $Fr^{-1}\approx9$. Plots are analogous to Figure \ref{fig:SingleRunPr1}.}\label{fig:SingleRunPrm8}
\end{figure*}


\begin{figure}
\includegraphics[width=\linewidth]{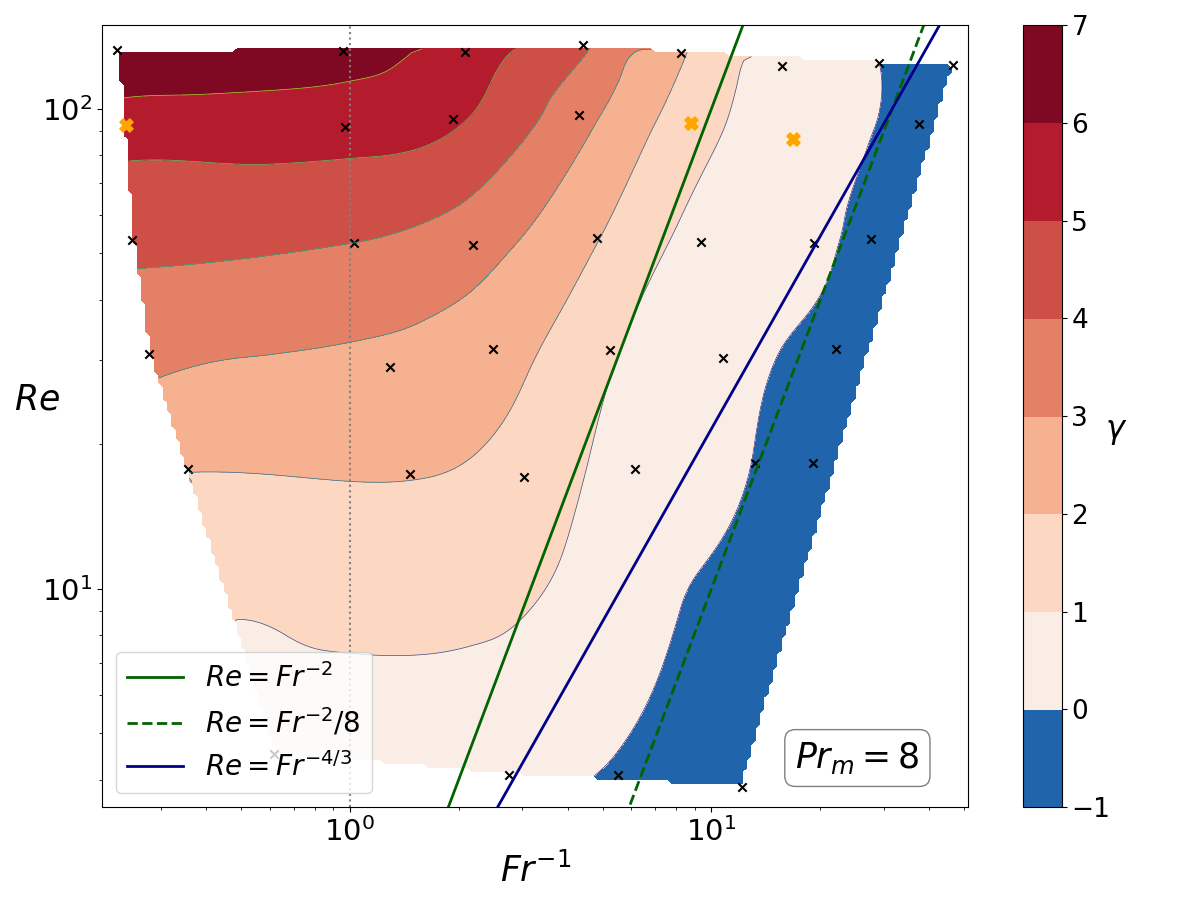}
\caption{Contour plot of the dynamo growth rate $\gamma$ in the space of $Re$ vs $Fr^{-1}$ for $Pr_m=8$ using sets 12-17 in Table \ref{tab:SimParam}. The blue and green lines are the scalings $Re^{3/4}=Fr^{-1}$ and $Re^{3/4}=Fr^{-3/2}$ corresponding to the wavenumber scalings $k_\nu=k_b$ and $k_\nu=k_O$, respectively. Black crosses are individual simulations and bold orange crosses correspond to simulations analyzed in Figure \ref{fig:SingleRunPrm8}. }\label{fig:ContourPlotPrm8}
\end{figure}

\begin{figure}
\includegraphics[width=\linewidth]{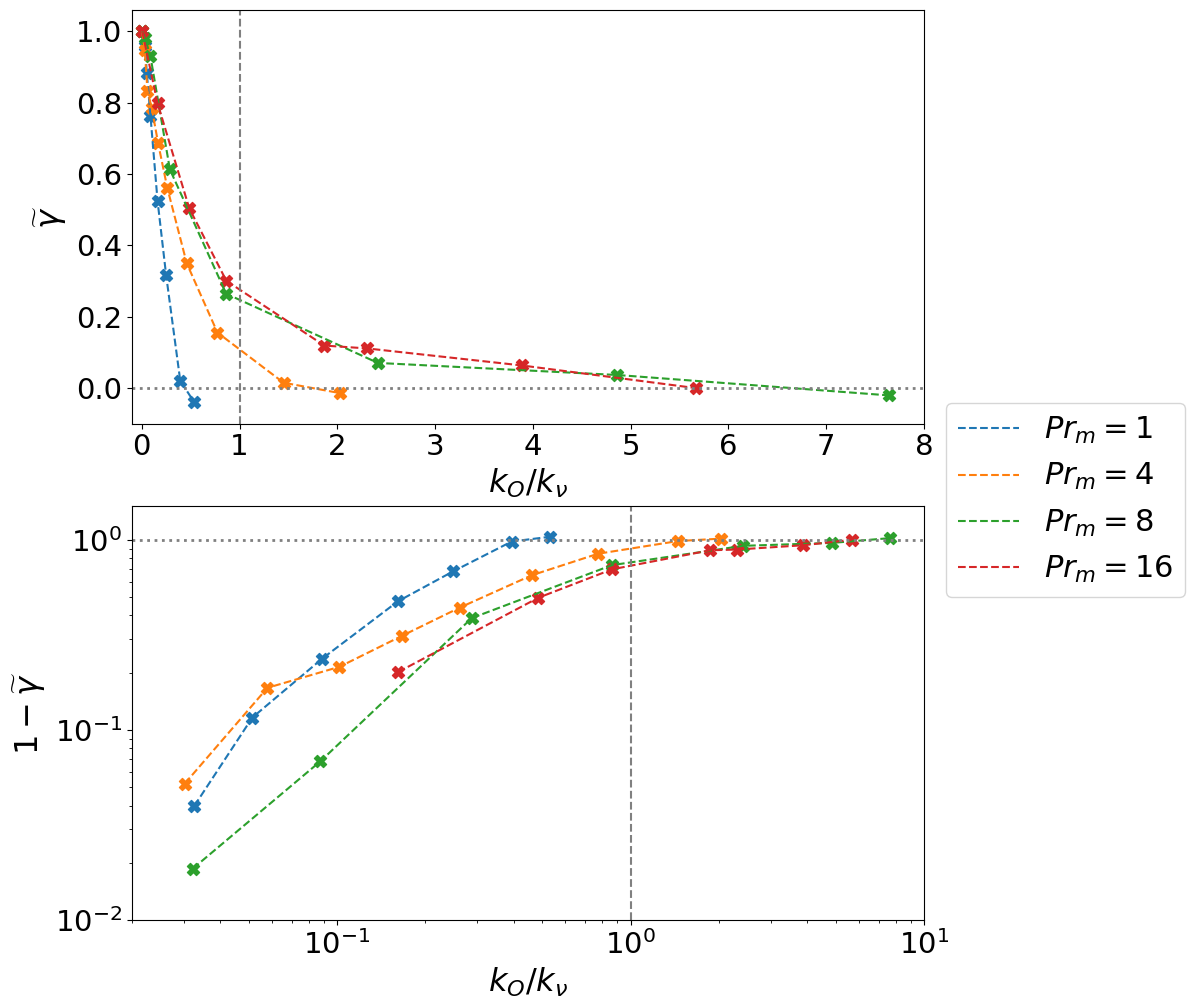}
\caption{Normalized growth rate $\widetilde \gamma$ versus $k_O/k_\nu$ at fixed $Re\approx 90$ for increasing $Pr_m\geq1$ using sets $6$, $11$, $16$, and $18$ in Table \ref{tab:SimParam}. The top panel plots $\widetilde \gamma$ directly while the bottom panel plots $1-\widetilde \gamma$ on a log-log scale. }\label{fig:NormalizedGammaPrm8}
\end{figure}

Figure \ref{fig:ContourPlotPrm1} shows that the $k_O\sim k_\nu$ scaling applies since the dashed green $m=2$ fit cleanly matches the onset boundary for $Re \gtrsim 30$, while any $m<2$ asymptote would be too shallow to match the boundary. The fit $Re^c=3.0Fr^{-2}$ corresponds to a critical buoyancy number $Rb^c=3$ at $Pr_m=1$. This implies that the $Pr_m=1$ dynamo will always be present in the stratified turbulence regime since, as discussed in Section \ref{sec:LengthScales}, $Rb\sim 1$ corresponds to the transition from the stratified turbulence regime to the VASF regime. 

\paragraph{Growth Rate Scaling} The adjacent contours to the left of the onset boundary appear to have equal slopes, implying that the scale separation between the Ozmidov and viscous scale also controls the scaling of the dynamo growth rate for $Rb>Rb^c$. We plot the normalized growth rate $\widetilde \gamma(Re,Rb)$ (defined in Section \ref{sec:OnsetCriterion}) at approximately fixed values of $Re$ versus $k_O/k_{\nu}=Rb^{-3/4}$ in Figure \ref{fig:Fig4_NormalizedGammaPr1}. The resulting set of curves all cross $\widetilde{\gamma}=0$ at approximately the same $k_O/k_{\nu}$, but $\widetilde{\gamma}(Re,Rb)$ still contains a modest $Re$ dependence. This may be because asymptotic values of $Re$ are only beginning to be reached at the highest available resolution. For reference, we superimpose empirical fits of the form
\begin{equation}\label{eq:EmpiricalFitPrm1}
    \widetilde{\gamma}=1-\left(\frac{Rb^{c}}{Rb}\right)^{n} 
\end{equation} 
and find that $n\approx0.5$ provides the most accurate fit at the highest accessible $Re$.

\begin{figure*}
\centering
\includegraphics[width=0.75\linewidth]{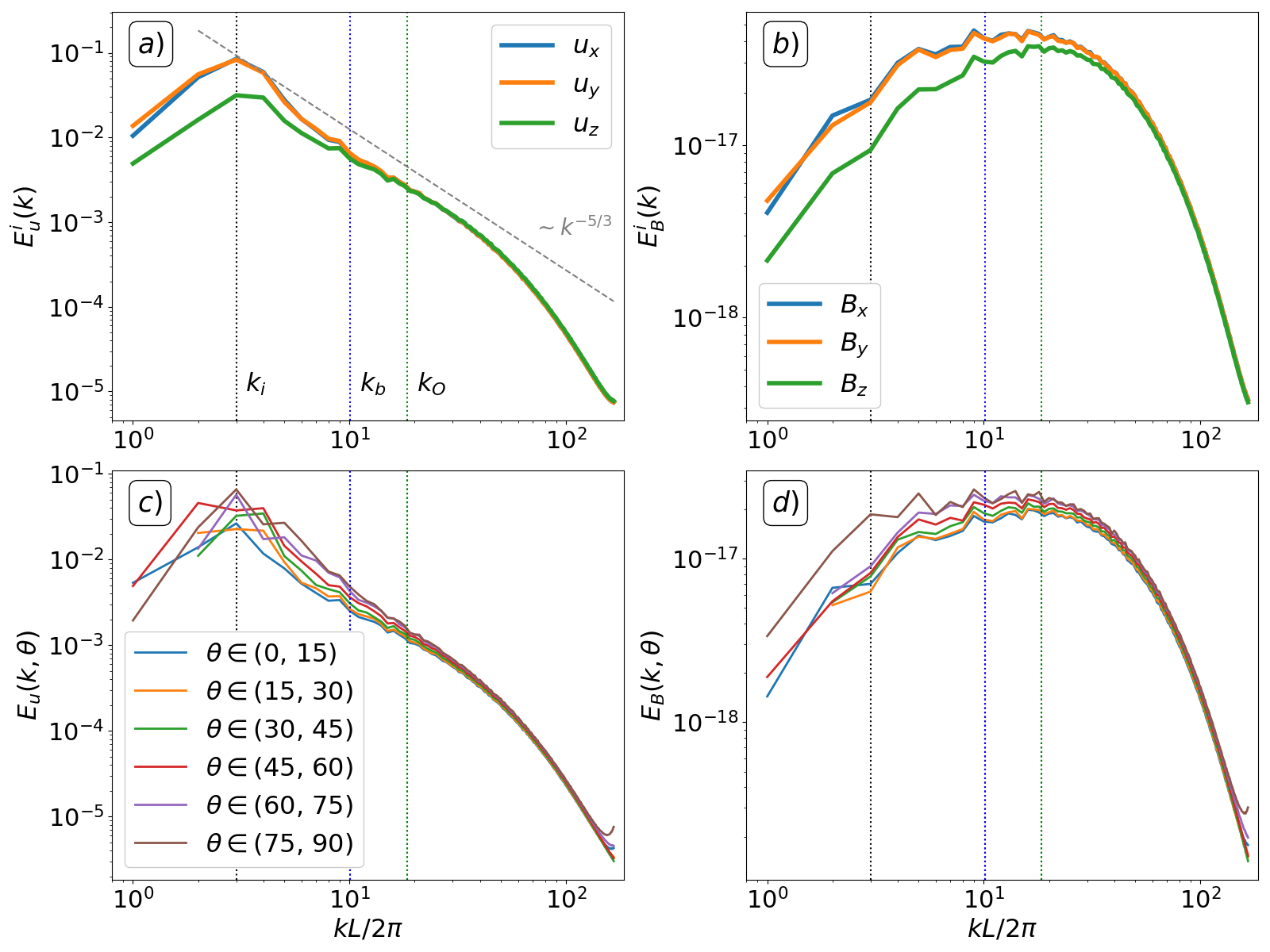}
\caption{  Spectral diagnostics of the $Pr_m=0.25$ simulation with strong stratification, $Re\approx360$, $Fr^{-1}\approx4$.  Plots are analogous to Figure \ref{fig:SingleRunPr1}.}  \label{fig:SingleRunLowPrm}
\end{figure*}

\subsubsection{High $Pr_m$ Regime}\label{sec:HighPrm}

We present the $Pr_m=8$ case in detail followed by a extension to $Pr_m=4$ and $Pr_m=16$.

\paragraph{Spectra Analysis} A representative single simulation shown in Figure \ref{fig:SingleRunPrm8} at $Pr_m=8$, $Re\approx90$ and strong stratification $Fr^{-1}=9$ ($Rb\approx1$) has similar but exaggerated characteristics compared to the $Pr_m=1$ case. The magnetic field is predominately horizontal, with the energy in the vertical component an order of magnitude smaller across all scales (Figure \ref{fig:SingleRunPrm8}b). The magnetic angular energy spectra anisotropy (Figure \ref{fig:SingleRunPrm8}d) is largest at small scales, but is progressively more isotropic at larger scales. This is consistent with the picture of viscous scales primarily driving the kinematic dynamo since the anisotropy at the viscous scales is quite high (Figure \ref{fig:SingleRunPrm8}c) for this strongly stratified case. We predict that as the dynamo saturates and the smallest eddies begin to feel feedback from the Lorentz force, the magnetic angular energy spectra anisotropy will spread to larger scales as larger and more anisotropic eddies take over driving the dynamo. 

Comparison with an unstratified case with $Fr^{-1}=0$ and a simulation in the VASF regime $Fr^{-1}=16$ ($Rb\approx0.4$) shows a pattern similar to the $Pr_m=1$ comparison in Figure \ref{fig:ComparisonPr1}. A comparison of the angular anisotropy $a(k)$ is likewise similar: at high $k$, $a_B(k)$ is relatively constant and increases alongside $a_u(k)$ with increasing stratification. This again supports the picture that the magnetic field anisotropy is controlled by the fluid anisotropy at the viscous scales.

\paragraph{Dynamo Onset and Scaling} We plot the growth rate contour for $Pr_m=8$ in Figure \ref{fig:ContourPlotPrm8}, which reveals that the onset curve $\gamma=0$ has shifted to the right compared to the $Pr_m=1$ case but still scales well with $Re\sim Fr^{-2}$. In other words, at dynamo onset, $k_O$ still scales with $k_\nu$ for $Pr_m=8$,  but $Rb^c$ has decreased to $Rb^c\approx 1/8$. The solid green line marks the transition between stratified turbulence to the left and the VASF regime to the right, and it is noticeable how the contour spacing sharply changes across the transition. This can be clearly seen in the plot of the normalized growth rates in Figure \ref{fig:NormalizedGammaPrm8} including other values of $Pr_m=1,4,8,16$ with fixed $Re\approx90$. The normalized growth rate curve indeed shifts to the right for increasing $Pr_m$, but across the transition point marked by the dashed vertical gray line, the growth rate curve seems to level out and decreases more slowly with $k_O/k_\nu$ for $k_O>k_\nu$. This highlights the importance of the $Rb=1$ transition, as well as hinting that the high-$Pr_m$ dynamo in the VASF regime could have a somewhat different character. Since $Rb$ controls the velocity field anisotropy at the viscous scales, it is plausible that $Rb^c$ becomes constant at higher $Pr_m$, as is already suggested by the $Rb^c(Pr_m)$ curve in Figure \ref{fig:Rbcrit}. However, simulations at even higher $Pr_m$ would be needed to confirm this.

\subsubsection{Low $Pr_m$ regime}\label{sec:LowPrm}

\begin{figure}[h]
\includegraphics[width=\linewidth]{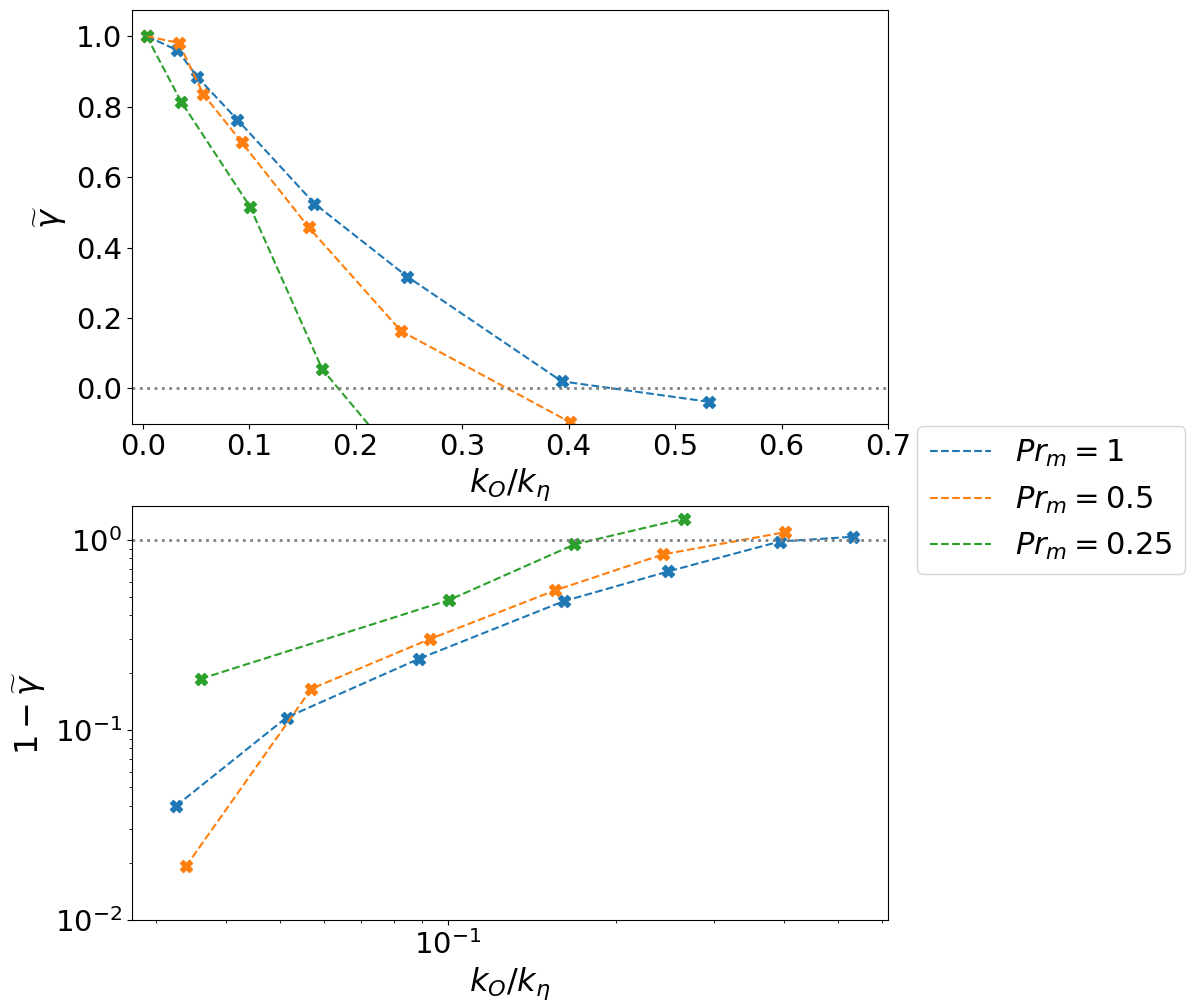}
\caption{Normalized growth rate $\widetilde \gamma$ versus $k_O/k_\eta$ at fixed $Rm=Pr_m Re\approx 90$ for decreasing $Pr_m\leq1$ using sets 1,2,6 in Table \ref{tab:SimParam}. Top plots $\widetilde \gamma$ directly while bottom plots $1-\widetilde \gamma$ on a log-log scale. }\label{fig:NormalizedGammaLowPrm}
\end{figure}

In the low-$Pr_m$ regime, the resistive scale moves into the inertial range ($k_\eta=Pr_m^{3/4}k_\nu$), and the dynamo is thought to be driven by a net dominance of stretching over diffusion by eddies with $k\lesssim k_\eta$, potentially in tandem with the forcing-scale eddies $k\sim k_i$ \citep{Iskakov2007}. If forcing scales do contribute, one might expect that $k_b$ or $k_{O}$ instead of $k_i$ would act as the largest dynamo-contributing eddy, since for larger scales anisotropy would likely cause diffusion to instead dominate over stretching. The dynamo would then shut off when the scale separation between $k_b$ or $k_O$ and $k_\eta$ became too small. This would correspond to a dynamo stability onset that scales as 
\begin{equation}
    Rm\sim Fr^{-m} \;\;\; (4/3\leq m\leq2),
\end{equation} 
with $m=4/3$ and $m=2$ implying $k_b\sim k_\eta$ and $k_O\sim k_\eta$ respectively.

To be truly in the low-$Pr_m$ regime, a simulation requires at least an order of magnitude separation between the resistive and viscous scales since $0.1<Pr_m<1$ corresponds to $k_\eta$ residing in the bottleneck region, and only for $Pr_m<0.1$ does the $k_\eta$ move into the inertial range. Additionally, achieving scale separation between the stratification scales is currently not possible, since even the highest-resolution unstratified simulations have marginal growth rates ($Rm\gtrsim Rm^c$). Nonetheless, we show simulation results for $Pr_m=0.5,0.25$, the limit of available resources. 

\paragraph{Single Simulation} Spectral diagnostics for a single simulation at $Re\approx360$, $Fr^{-1}=4$, and $Pr_m=0.25$ are shown in Figure \ref{fig:SingleRunLowPrm}. The magnetic field anisotropy has a pattern opposite to the high $Pr_m$ case in Figure \ref{fig:SingleRunPrm8}. The magnetic field is primarily horizontal  for the larger scales, but becomes isotropic at smaller scales (Figure \ref{fig:SingleRunLowPrm}b). Likewise, the angular energy spectra are anisotropic at larger scales and more isotropic at smaller scales (Figure \ref{fig:SingleRunLowPrm}d). This supports the picture that the fluid eddies at the  (now larger) resistive scale primarily contribute to the dynamo and set the magnetic field anisotropy. Magnetic fields with $k>k_\eta$ are simply dissipated and lose their anisotropy. 

\paragraph{Onset Criterion and Growth Rate Scaling} We plot the normalized growth rate for $Pr_m=1,0.5,0.25$ in Figure \ref{fig:NormalizedGammaLowPrm}. The green curve for $Pr_m=0.25$ shows that the dynamo shuts off even earlier than in the $Pr_m=1$ case, suggesting that the $k_b\sim k_\eta$ scaling is unlikely. Instead, it seems that the $k_O\sim k_\eta$ scaling applies, but with a critical scale separation between $k_O $ and $k_\eta$ needed to enable dynamo growth increasing with decreasing $Pr_m$ (see Figure \ref{fig:Rbcrit}). In other words, the $Pr_m=0.25$ dynamo requires a larger critical $Rb_m^{c}\approx 9$ for the dynamo to operate than for $Pr_m=1$ where we had $Rb_m^{c}=Rb^{c}\approx3$. The $Rb_m^{c}(Pr_m)$ curve may similarly qualitatively follow the critical magnetic Reynolds $Rm^c(Pr_m)$ curve, which increases for $Pr_m\lesssim1$, peaks around $Pr_m\simeq \!0.1$ when $k_\eta$ resides in the bottleneck region, and then decreases and plateaus to a constant for $Pr_m\leq 0.1$, when $k_\eta$ enters the inertial range. This suggests that the measured value of $Rb_m^c\approx9$ at $Pr_m=0.25$ could be nearing an upper bound for the asymptotic value of $Rb_m^c$ for $Pr_m\ll1$, although larger simulations would be needed to confirm this.

\section{Application to Stellar Radiative Zones}\label{sec:ApplicationToTC}
In order to determine the existence of the SSD in stellar radiative zones, we need representative parameter values of $Re$, $Fr$, and $Pr_m$. We turn to the solar tachocline, where helioseismology and solar models have provided relatively precise parameter estimates. Stratified turbulence in the upper region of the solar tachocline could potentially be driven by a combination of horizontal/vertical shear turbulence and convective overshoot \citep{zahn1992circulation,miesch2005large}. The horizontal and vertical shears arise from latitudinal and radial solar differential rotation, respectively.

We consider the case of driving due to horizontal shear turbulence since in the tachocline the vertical shear is thought to be stable \citep{schatzman2000shear}, although this is not generally true in stellar radiative zones (see \citet{heger2000presupernova} and references within). The turbulent velocity around the mean horizontal shear flow can be taken as $U=\widetilde{U}\cdot100\text{m/s}$ \citep{miesch2005large,cope2020dynamics}. The horizontal integral scale is usually taken as $l_i=\widetilde{l}_i R_\odot$ \citep{zahn1992circulation,cope2020dynamics} and the Brunt V\"ais\"al\"a frequency to be on the order of a millihertz \citep{hughes2007solar}, $N=\widetilde{N}\cdot 1\text{mHz}$. With $\nu=\widetilde{\nu}\cdot3\times10^{-3}\text{m}^2\text{/s} $ \citep{hughes2007solar}, the resulting Reynolds number is $Re=10^{13}\cdot\widetilde{U}\widetilde{l_i}/\widetilde{\nu}$ and the Froude number is $Fr= 10^{-4}\cdot \widetilde{U}/(\widetilde{N}\widetilde{l_i})$. Estimates for the magnetic Prandtl number at the tacholine place $10^{-3}\leq Pr_m\leq 7\times10^{-2}$ which we write as $Pr_m=10^{-2}\cdot \alpha$ \citep{hughes2007solar}. Using these, we calculate the stratification and dissipation length scales in Table \ref{tab:LengthScales}.

\begin{table}[h]
\setlength{\tabcolsep}{10pt} 
\renewcommand{\arraystretch}{2.25} 
    \centering
    \begin{tabular}{c|c}
         $l_b$    &  $\left(\frac{\widetilde{U}}{\widetilde{N}}\right)10^5m$\\
    \hline
         $l_O$    &  $\left(\frac{\widetilde{U}^{3/2}}{\widetilde{N}^{3/2}\widetilde{l_i}^{1/2}}\right)10^3m$\\
    \hline
        $l_\eta$ & $\left(\frac{\widetilde{l_i}^{1/4}\widetilde{\nu}^{3/4}}{\alpha^{3/4}\widetilde{U}^{3/4}}\right)$2m \\
    \hline
        $l_\nu$  & $\left(\frac{\widetilde{l_i}^{1/4}\widetilde{\nu}^{3/4}}{\widetilde{U}^{3/4}}\right)0.07m$ 
    \end{tabular}
    \caption{Length scales in the tachocline.}
    \label{tab:LengthScales}
\end{table}

The scale separation between the Ozmidov and resistive scales $l_O/l_\eta\approx600$ is more than two orders of magnitude and corresponds to a magnetic buoyancy Reynolds number of 
\begin{equation}
    Rb_m\approx5\times 10^3\left(\frac{\widetilde{U}^{3}\alpha}{\widetilde N^{2}\widetilde{l}_i\widetilde{\nu}}\right).
    \label{eq:TCRbm}
\end{equation}
If we take the value $Rb_m^{c}=O(10)$ at $Pr_m=0.25$ as a rough upper bound for $Rb_m^{c}(Pr_m\ll 1)$ as argued in Section \ref{sec:LowPrm}, then Equation \eqref{eq:TCRbm} plausibly predicts an active small-scale dynamo in the parameter regime of the solar tachocline.

Equation \eqref{eq:TCRbm} is fairly sensitive to parameter estimates, and it is clear that the strength of the small-scale dynamo may have strong vertical variation across the tachocline. For example, near the top of the tachocline, $\widetilde N<<1$ (since $N=0$ at the convective-radiative interface) and $l_i$ could instead be argued to be on the order of a convective plume ($l_i\sim H_p<<R_\odot$, where $H_p\sim0.05R_\odot$ is a pressure scale height), significantly increasing $Rb_m$ due to both effects. Additionally, in radiative zones with radial differential rotation unstable to a vertical shear instability, $l_i$ would be on the order of the vertical shear gradient length scale, which leads to a much larger $Rb_m$ compared to the driving by a horizontal shear instability considered here. On the other hand, the true size of $U$ is poorly understood and may be lower than its upper bound (as well as variation of the driving mechanisms with height) which would easily lead to a reduction of $Rb_m$ due to the sensitive scaling $Rb_m\sim\widetilde U^3$.

\subsection{Qualifications and Discussion} 

\paragraph{Role of Other Instabilities} In reality, stellar radiative zones are affected by a variety of different instabilities that will saturate nonlinearly in complex ways. These may affect the SSD, and vice versa. Understanding the saturated state would require reexamining these instabilities with the SSD in mind, which is outside the scope of this paper. For example, a growing subequipartition toroidal field in the tachocline (e.g. as a result of a large-scale dynamo) would be accompanied by the much faster growing SSD, and the final state would be a mixture of the saturated large-scale dynamo and SSD. Indeed, in addition to fossil fields, the SSD may provide the seed magnetic fluctuations needed for the Tayler-Spruit or mean-field dynamos in radiative zones. 

\paragraph{Low Magnetic Prandtl Number} The existence of the low $Pr_m$ SSD below $Pr_m=O(10^{-2})$ is a core assumption for the validity of our proposed scalings and predictions for the SSD in stellar radiative zones.  Computational resources limit the maximum value of $Rm$ which is currently too close to the large value of the critical $Rm$ and leads to small growth rates. However, the conclusions of \cite{Iskakov2007,Schekochihin2007} indicate that, at the extremely high $Rm$ in stars, the low $Pr_m$ SSD should be well within the unstable regime when considering isotropic, homogeneous turbulence. The results of this paper build on and extend this prediction to stably stratified turbulence.

\paragraph{Low Thermal Prandtl Number} The above estimate also does not consider the effect of a low $Pr$ as discussed in Section \ref{sec:ThermalPr}. Using Table \ref{tab:LengthScales}, the buoyancy number is $Rb=(l_O/l_\nu)^{4/3}\sim 5\times 10^5$ subject to the same caveats as Equation \eqref{eq:TCRbm}. The low $Pr\sim 10^{-6}$ in the tachocline marginally satisfies the criterion $Pr\leq Rb^{-1}$ (Equation \eqref{eq:LowPr}) implying that the tachocline likely contains a stronger SSD than suggested by the prediction based on $Pr=1$ following Equation \eqref{eq:TCRbm}.

\paragraph{Mean Shear} Additionally, the effects of a horizontal or a vertical mean shear on the SSD are not considered in this paper. The effect of mean shear on the SSD has only been directly studied in the unstratified case, where full solutions of the Navier-Stokes equation for the shear flow show that the turbulence resulting from shear instabilities helps drive the SSD \citep{singh2017enhancement,tobiasShear2019}, while prescribed flows at much higher $Rm$ show a suppression of the SSD \citep{tobias2013shear}. The significant complexity added when combining shear and stratification makes it difficult to estimate whether shear even decreases or increases SSD action and is left for future study.

\section{Summary and Conclusion}\label{sec:conc}
We present theoretical arguments and simulations of the kinematic small-scale dynamo in stably stratified turbulence to determine the dynamo onset criterion, study the scaling of the dynamo growth rate with increasing stratification, and characterize the dynamo-generated magnetic field. All simulations solve the MHD Boussinesq equations using the SNOOPY code with isotropic, time-correlated forcing and $Pr=1$. The main results are itemized below:

\begin{itemize}
    \item In the presence of stratification with $Pr\gtrsim1$ and $Rm>Rm^c$, direct numerical simulations suggest that the additional criterion for the onset of the SSD is $Rb>Rb^c$ for $Pr_m\geq 1$ and $Rb_m=Pr_m Rb>Rb_m^c$ for $Pr_m\leq1$, where $Rb=ReFr^{-2}$ and $Rb_m$ are the buoyancy Reynolds and magnetic buoyancy Reynolds numbers. $Rb_m^c$ and $Rb^c$ are both dependent on $Pr_m$, analogous to $Rm^c$. Simulations and theoretical arguments suggest that $Rb_m^c\simeq 9$ is a likely upper bound for $Rb_m^c$ in the low $Pr_m$ limit, while $Rb^c\simeq 0.1$ for the high $Pr_m$ limit. For $Pr_m=1$, $Rb_m^c=Rb^c\approx3$.

    \item The SSD onset criterion is satisfied in the the solar tachocline with $Rb_m=O(10^3)$, assuming $Rb_m^c=O(10)$ for $Pr_m\ll 1$. However, we also argue that the low thermal Prandtl number of the tachocline softens the onset criterion. Therefore, the results imply that a SSD is plausibly active in the tachocline provided a combination of horizontal/vertical shear turbulence and/or convective overshoot serves as a driving mechanism for stratified turbulence. 
    
    \item Analyzing individual simulations shows that anisotropy in both of the components of the magnetic field and the angular energy spectrum is roughly constant across all scales and is primarily set by the anisotropy present at the viscous/resistive scales for high/low $Pr_m$.  Vertical modes ($\textbf{k}\parallel \textbf{g}$) of the magnetic field contain more energy than horizontal modes in the angular energy spectrum, and vertical components of the magnetic field contain less energy than the horizontal components across all scales. This is unlike the velocity field, which is out of equipartition only for scales $k<k_b$ and whose anisotropy varies strongly with scale. 
\end{itemize}

The presence of a small-scale dynamo in differentially rotating regions of radiative zones could have important effects. When the SSD saturates, it will likely reach approximate equipartition with at least the energy available in the isotropic fluid scales $k\geq k_O$ which would allow feedback on the flow through Maxwell stresses and/or affect any possible large-scale dynamo mechanism. As an example of the latter, the magnetic fluctuations in a radial shear flow would satisfy conditions for possible operation of the magnetic-shear current effect \citep{squirePRL_MSC}, allowing a large-scale toroidal field to grow and be directly stored in the stratified region. The interplay of such effects with the saturated state of the stably stratified SSD will be the subject of future work.

\acknowledgments
We would like to thank Daniel Lecoanet, Matthew Kunz, Adam Burrows, Alexander Philippov, and Ammar Hakim for insightful comments and discussions throughout this work. Simulations were carried out on the Perseus and Eddy clusters at Princeton University. V. S. was supported by Max-Planck/Princeton Center for Plasma Physics (NSF grant PHY-1804048). A. B. was supported by the DOE Grant for the Max Planck Princeton Center (MPPC). J. S. was supported by a Rutherford Discovery Fellowship RDF-U001804 and Marsden Fund grant UOO1727, which are managed through the Royal Society Te Ap\=arangi.

\appendix
\section{Scaling Analysis}\label{sec:ScalingAnalysis}
Here we present a scaling analysis of the governing equations for an alternative but closely related perspective on stratified turbulence. Scaling of the Boussinesq equations in Section \ref{sec:ScalingAnalysisBoussinesq} serves as a plausible derivation of the two stratified turbulence regimes and helps with understanding the nature of the corresponding velocity fields. Section \ref{sec:ScalingAnalysisInduction} then extends the scaling assumptions to the induction equations and makes predictions on when and why the dynamo shuts off with increasing stratification. 

\subsection{Boussinesq Equations}\label{sec:ScalingAnalysisBoussinesq}
In the Boussinesq approximation with gravity $\textbf{g}=g\hat{z}$ and background density profile $\overline{\rho}(z)$ relative to a reference density $\rho_0$, the perturbative velocity $\textbf{u}'$, density $\rho'$, and pressure $p'$ satisfy
\begin{equation}\label{eq:BousMomentum}
    \partial_t \textbf{u}'+\textbf{u}'\cdot \nabla' \textbf{u}'=-\frac{1}{\rho_0}\nabla p'-\frac{\rho' g}{\rho_0}\hat{z}+\nu {\nabla'}^2\textbf{u}',
\end{equation}
\begin{equation}\label{eq:BousDen}
    \partial_{t'}\rho'+\textbf{u}'\cdot \nabla' \rho'=-\frac{d\overline{\rho}}{dz}u_z'-+\kappa {\nabla'}^2\rho',
\end{equation}
\begin{equation}\label{eq:BousIncomp}
    \nabla\cdot \textbf{u}'=0,
\end{equation}
where $\nu$ is the viscosity and $\kappa$ is the thermal diffusivity. Temperature perturbations $\rho'/\rho_0=-\alpha_p T'$ are directly related to density perturbations through the isobaric thermal expansion coefficient $\alpha_p$. The dimensional variables here are labeled with primes, while the dimensionless variables in the scaling analysis will be left unprimed. 

Due to the buoyant restoring force, vertical displacements in Eqs \eqref{eq:BousMomentum}-\eqref{eq:BousIncomp} undergo oscillations at the Brunt-V\"ais\"al\"a frequency $N^2=-g(d\overline{\rho}/dz)/\rho_0>0$. Consider a horizontal velocity scale $U$ and horizontal length scale $l_h$ imposed on the system. The dimensionless measure of stratification is the Froude number $Fr=U/(l_h N)$, comparing buoyancy timesscales $N^{-1}$ to advection time scales $l_h/U$.  \cite{billant2001self,godoy2004vertical} provide dominant balance arguments for Eqs \eqref{eq:BousMomentum}-\eqref{eq:BousIncomp} under strong stratification $Fr\ll1$ resulting in the dimensionalized quantities below:

\begin{equation}
\textbf{u}_h'=U \textbf{u}_h,\;u_z'=U\frac{Fr^2}{\alpha}u_z,\;\rho'=\frac{\rho_0U^2}{gl_v}\rho,\;p=\rho_0U^2 p',
\end{equation}

\begin{equation}
x'=l_h x,\;y'=l_h y,\;z'=l_v z,\;t'=\frac{l_h}{U} t,
\end{equation}
where $l_v=\alpha l_h$ is the emergent typical vertical scale of the flow. The above scalings lead to the following dimensionless equations generally describing strong Boussinesq stratification:

\begin{equation}\label{eq:GeneralDimlessHorzMom}
    \frac{D_h\textbf{u}_h}{Dt}+\frac{Fr^2}{\alpha^2}u_z \nabla_z \textbf{u}_h=-\nabla_h p+\frac{1}{Re}(\nabla_h^2+\frac{1}{\alpha^2}\nabla_z^2)\textbf{u}_h,
\end{equation}

\begin{equation}\label{eq:GeneralDimlessVertMom}
     Fr^2(\frac{D_h u_z}{Dt}+\frac{Fr^2}{\alpha^2}u_z \nabla_z u_z)=-\nabla_z p-\rho+\frac{Fr^2}{Re}(\nabla_h^2+\frac{1}{\alpha^2}\nabla_z^2)u_z,
\end{equation}
\begin{equation}
    \frac{D_h\rho}{Dt}+\frac{Fr^2}{\alpha^2}u_z\nabla_z\rho=u_z+\frac{1}{RePr}(\nabla_h^2+\frac{1}{\alpha^2}\nabla_z^2)\rho,
\end{equation}
\begin{equation}
    \nabla_h\cdot \textbf{u}_h+\frac{Fr^2}{\alpha^2}\nabla_z u_z=0,
\end{equation}
where $D/Dt=\partial_t+u\cdot\nabla$ is the convective derivative, the Reynolds number is $Re=Ul_h/\nu$, the Prandtl number is $Pr=\nu/\kappa$, and the subscripts $h$ and $z$ correspond to horizontal and vertical components. The horizontal momentum Equation \eqref{eq:GeneralDimlessHorzMom} contains two possible balances depending on the buoyancy Reynolds number $Rb=Re Fr^2$ which measures the size of the vertical advection to the vertical diffusion term. When $Rb\gg1$, the diffusion terms can be dropped, and dominant balance sets $\alpha=Fr$ resulting in the following system of equations:
\begin{equation}\label{eq:StrongStratHorzMom}
    \frac{D_h\textbf{u}_h}{Dt}+u_z\cdot \nabla_z \textbf{u}_h=-\nabla_h p,
\end{equation}

\begin{equation}\label{eq:StrongStratVerMom}
     0=-\nabla_z p-\rho,
\end{equation}
\begin{equation}
    \frac{D_h\rho}{Dt}+u_z\nabla_z\rho=u_z,
\end{equation}
\begin{equation}\label{eq:VertDiv}
    \nabla_h\cdot \textbf{u}_h+\nabla_z u_z=0.
\end{equation}

The limit of strong stratification $Fr\ll1$ with $Rb\gg1$ thus leads to neglecting the vertical acceleration term in Equation \eqref{eq:StrongStratVerMom} while the vertical advection term in Equation \eqref{eq:StrongStratHorzMom} and vertical divergence in Equation \eqref{eq:VertDiv} stay order one. Although the vertical velocity is small $u_z'\sim FrU$, the vertical velocity length scales do not collapse to zero and instead are restricted to the buoyancy scale $l_v\sim l_b=U/N$ (independent of $Re$), leading to balance of horizontal and vertical gradients. These equations support internal gravity waves and smaller scale 3D turbulent-like structures at and below the buoyancy scale, both of which are observed in simulations \citep{lindborg_2006,brethouwer_billant_lindborg_chomaz_2007,Waite2011}. This is the stratified turbulence regime discussed in Section \ref{sec:LengthScales} with the scale separation requirement $k_O\gg k_\nu$ that is equivalent to $Rb\gg1$. 

On the other hand, for $Rb<1$ the vertical diffusion terms dominate the vertical advection terms (assuming $Pr\gtrsim 1$) and dominant balance of the vertical diffusion term in the horizontal momentum Equation \eqref{eq:GeneralDimlessHorzMom} sets $\alpha^2 Re=1$ \citep{godoy2004vertical}. Vertical scales then become negligible, $l_v=Re^{-1/2}l_h$ (independent of $Fr$). The resulting equation set,

\begin{equation}
    \frac{D_h\textbf{u}_h}{Dt}=-\nabla_h p+\nabla_z^2\textbf{u}_h,
\end{equation}

\begin{equation}
     0=-\nabla_z p-\rho,
\end{equation}
\begin{equation}
    \frac{D_h\rho}{Dt}=u_z+\frac{1}{Pr}\nabla_z^2\rho,
\end{equation}
\begin{equation}
    \nabla_h\cdot \textbf{u}_h=0,
\end{equation}
physically represents vertically, viciously coupled quasi-2D planes of flow. Indeed, simulations are typically characterized by thin, large-scale, stable horizontal layers that are missing smaller-scale features due to the suppression of instabilities and transition to turbulence by viscosity \citep{brethouwer_billant_lindborg_chomaz_2007}. This is the viscosity-affected stratified flow regime (VASF) discussed in Section \ref{sec:LengthScales} with the scale separation requirement $k_O<k_\nu$ that is equivalent to $Rb<1$. 

\subsection{Induction Equation}\label{sec:ScalingAnalysisInduction}
\begin{figure}
    \centering
    \includegraphics[width=\linewidth]{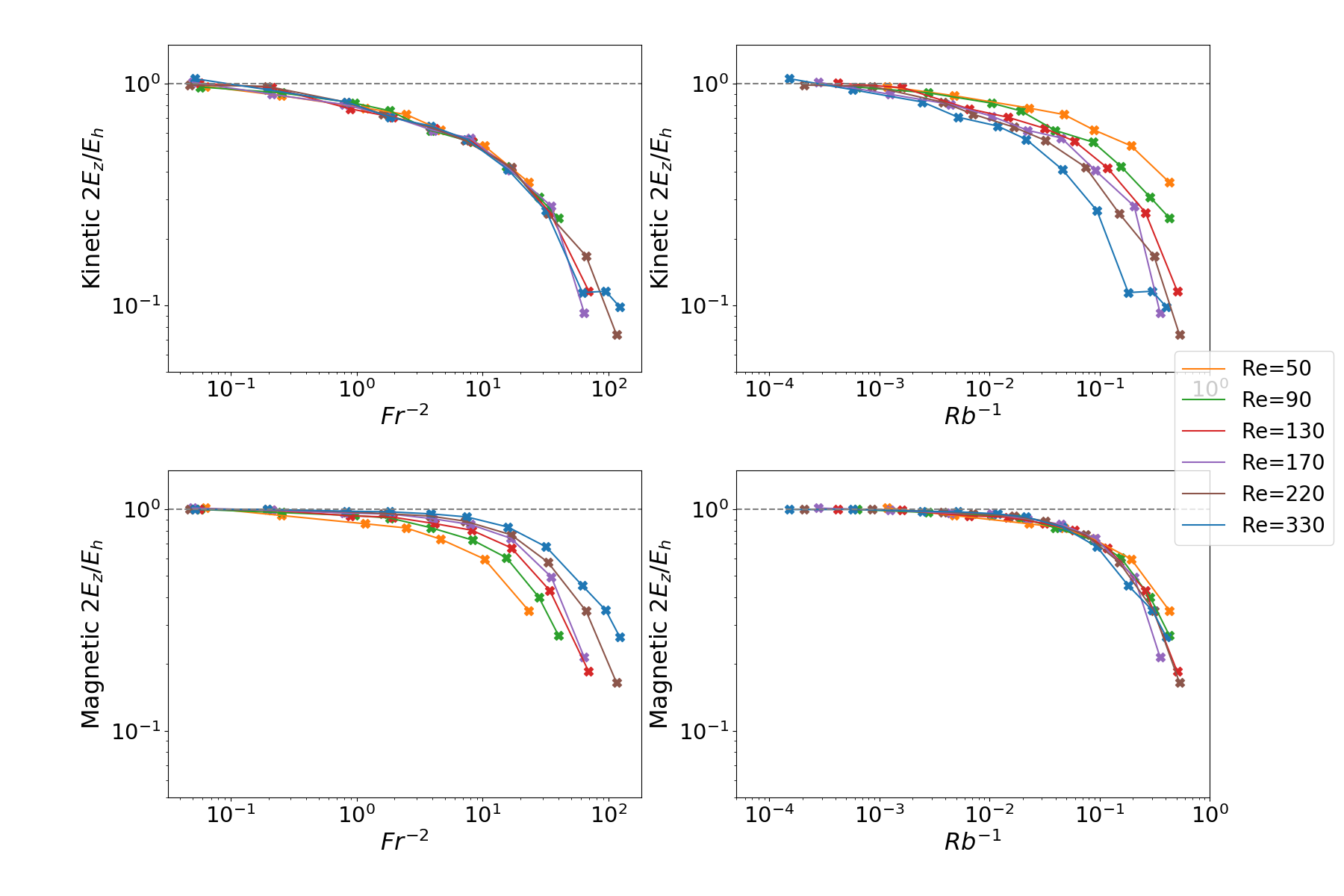}
    \caption{Scaling of vertical to horizontal $2E_z/E_h$ energy of the velocity and magnetic fields versus $Fr^{-2}$ and $Rb$ for $Pr_m=1$. We use the ratio of the vertical to horizontal energies as a proxy for the ratio of the vertical to horizontal field magnitudes (e.g. $2E_z/E_h\sim2B_z^2/B_h^2$ for the magnetic fields). In the isotropic case, $2E_z/E_h=1$ for both the kinetic and magnetic energies. }
    \label{fig:ScalingAmplitudes}
\end{figure}

We extend the scaling analysis of Section \ref{sec:ScalingAnalysis} for insight into the dynamo behavior in the presence of strong stratification. We assume the magnetic field components scale in the same way as the velocity field, $B_z=(Fr^2/\alpha) B_h$, and likewise assume that the magnetic fields vary on similar horizontal $l_h$ and vertical $l_v=\alpha l_h$ length scales. When comparing with simulations, these assumptions are seen to be incorrect: $B_z/B_h$ scales with $Rb$ in the stratified turbulence regime (when $\alpha=Fr$), while $u_z/u_h$ scales with $Fr$; see Figure~\ref{fig:ScalingAmplitudes}. We suspect this behavior occurs because the magnetic field anisotropy is primarily set by the anisotropy $a_u$ of the viscous/resistive scale eddies in the high/low $Pr_m$ regime (see Section \ref{sec:Results}), while $a_u$ is determined by $Rb$ instead of $Fr$. Despite this minor discrepancy, the scaling analysis provides valuable qualitative insight; it correctly predicts that the dynamo onset criteria scale with $Rb$ in the stratified turbulence regime and that the dynamo is killed near the transition to the VASF regime.

Application of the scaling assumptions to the induction equation gives

\begin{align}
\begin{split}
\frac{D_h}{Dt}\textbf{B}_h+\frac{Fr^2}{\alpha^2}u_z\nabla_z \textbf{B}_h=(\textbf{B}_h\nabla_h+\frac{Fr^2}{\alpha^2}B_z\nabla_z)\textbf{u}_h\\+\frac{1}{Re Pr_m}(\nabla_h^2+\frac{1}{\alpha^2}\nabla_z^2)\textbf{B}_h,
\end{split}
\end{align}

\begin{equation}
\begin{split}
\frac{D_h}{Dt}B_z+\frac{Fr^2}{\alpha^2}u_z\nabla_z B_z=(\textbf{B}_h\nabla_h+\frac{Fr^2}{\alpha^2}B_z\nabla_z)u_z\\+\frac{1}{Re Pr_m}(\nabla_h^2+\frac{1}{\alpha^2}\nabla_z^2)B_z,
\end{split}
\end{equation}

\begin{equation}
    \nabla_h \cdot \textbf{B}_h+\frac{Fr^2}{\alpha^2}\nabla_z B_z=0.
\end{equation}
Consider first the strongly stratified turbulence limit $Rb>>1$.  With $\alpha=Fr$ constrained from the momentum equation, the induction equation takes on the form 

\begin{equation}
\frac{D_h}{Dt}\textbf{B}_h+u_z\nabla_z \textbf{B}_h=(\textbf{B}_h\nabla_h+B_z\nabla_z)\textbf{u}_h+\frac{1}{Rb Pr_m}\nabla_z^2\textbf{B}_h,
\end{equation}

\begin{equation}
\frac{D_h}{Dt}B_z+u_z\nabla_z B_z=(\textbf{B}_h\nabla_h+B_z\nabla_z)u_z+\frac{1}{Rb Pr_m}\nabla_z^2B_z,
\end{equation}
\begin{equation}
    \nabla_h \cdot \textbf{B}_h+\nabla_z B_z=0.
\end{equation}
This corresponds to the usual form of the isotropic induction equation but with a lower "effective" magnetic Reynolds number $RbPr_m$ as well as an anisotropic resistivity. We have defined $Rb_m=RbPr_m$ as the magnetic buoyancy Reynolds number. Taken at face value, it suggests a dynamo should be possible if $Rb_m$ is larger than a critical $Rb^{c}_m$ analogous to the typical requirement $Rm>Rm^c$.

On the other hand, in the VASF regime ($Rb<1$), the vertical advection and vertical divergence terms drop out, giving the equation set (with $\alpha=Re^{-1/2}$). 

\begin{equation}
\frac{D_h}{Dt}\textbf{B}_h=(\textbf{B}_h\nabla_h)\textbf{u}_h+\frac{1}{Pr_m}\nabla_z^2\textbf{B}_h,
\end{equation}

\begin{equation}
\frac{D_h}{Dt}B_z=(\textbf{B}_h\nabla_h)u_z+\frac{1}{Pr_m}\nabla_z^2B_z,
\end{equation}

\begin{equation}
    \nabla_h \cdot \textbf{B}_h=0,
\end{equation}
which decouples the horizontal and vertical components of the induction equation, implying that no dynamo can be possible. Note that the scaling result is independent of $Pr_m$. A higher $Pr_m$ would only lead to a slower resistive decay of magnetic energy.

%

\vspace{5mm}


\software{SNOOPY \citep{SNOOPY2005A&A}}





\bibliography{main}{}
\bibliographystyle{aasjournal}



\end{document}